\documentclass{nsr}
\usepackage{amsmath,graphicx,array}
\usepackage{dcolumn,soul}%

\usepackage{amsthm}
\usepackage[figuresright]{rotating}%
\usepackage{algorithm, algorithmicx, algpseudocode, multirow, booktabs, threeparttable}
\usepackage{listings}%
\usepackage{hyperref}

\makeatletter
\def\uns{\ifmmode\,\else$\,$\fi}%

\makeatother

\jvol{XX}
\jnum{X}
\jyear{Year}
\doi{10.1093/nsr/XXXX}
\received{XX XX Year}
\revised{XX XX Year}
\accepted{XX XX Year}

\begin{document}

\dhead{RESEARCH ARTICLE}

\subhead{Information Science}

\title{Canine EEG Helps Human: Cross-Species and Cross-Modality Epileptic Seizure Detection via Multi-Space Alignment}

\author{Ziwei Wang$^{1,2}$}

\author{Siyang Li$^{1,2}$}

\author{Dongrui Wu$^{1,2,*}$}

\affil{$^1$Ministry of Education Key Laboratory of Image Processing and Intelligent Control, School of Artificial Intelligence and Automation, Huazhong University of Science and Technology, Wuhan 430074, China}
\affil{$^2$Hubei Key Laboratory of Brain-inspired Intelligent Systems, School of Artificial Intelligence and Automation, Huazhong University of Science and Technology, Wuhan 430074, China}

\authornote{\textbf{Corresponding author.} Email: drwu09@gmail.com}

\abstract[ABSTRACT]{Epilepsy significantly impacts global health, affecting about 65 million people worldwide, along with various animal species. The diagnostic processes of epilepsy are often hindered by the transient and unpredictable nature of seizures. Here we propose a multi-space alignment approach based on cross-species and cross-modality electroencephalogram (EEG) data to enhance the detection capabilities and understanding of epileptic seizures. By employing deep learning techniques, including domain adaptation and knowledge distillation, our framework aligns cross-species and cross-modality EEG signals to enhance the detection capability beyond traditional within-species and within-modality models. Experiments on multiple surfaces and intracranial EEG datasets of humans and canines demonstrated substantial improvements in detection accuracy, achieving over 90\% AUC scores for cross-species and cross-modality seizure detection with extremely limited labeled data from the target species/modality. To our knowledge, this is the first study that demonstrates the effectiveness of integrating heterogeneous data from different species and modalities to improve EEG-based seizure detection performance. This is a pilot study that provides insights into the challenges and potential of multi-species and multi-modality data integration, offering an effective solution for future work to collect huge EEG data to train large brain models.}

%\jelcode{Pa, J6, P16, E22}

\keywords{Electroencephalogram, Automatic Seizure Detection, Transfer Learning, Domain Adaptation, Knowledge Distillation}

\maketitle

\section{INTRODUCTION}\label{sec:intro}
Epilepsy is a chronic disorder characterized by sudden abnormal neuronal discharges, resulting in transient brain dysfunction\cite{wang2021electric}. It affects approximately 1\% of the population (65 million) worldwide, including adults, infants, and young children, and is highly prevalent among various animal species as well\cite{bui2018dentate,wang2021electric,karoly2021cycles}. Epileptic individuals exhibit a range of symptoms\cite{Tellez2007long}, including generalized convulsions, loss of consciousness, debilitation, and recurrent seizures. These symptoms may cause irreversible brain damage and life-threatening situations\cite{kerr2012impact}, which can lead to employment restrictions and social isolation.

Consequently, early diagnosis and preventative measures for epilepsy are of paramount importance. Sophisticated medical imaging modalities such as computed tomography and magnetic resonance imaging enable epilepsy detection by identifying lesions and providing comprehensive spatial information\cite{Wang2023automated}. However, these techniques lack temporal resolution and cannot capture ongoing seizures. To overcome these limitations and facilitate timely epilepsy diagnosis, electroencephalogram (EEG) signals are utilized for their high temporal resolution.

EEG provides a diagnostic test that detects epileptiform discharges by monitoring voltage fluctuations caused by neural activities of the brain\cite{saab2020weak}. EEG can be further categorized into scalp EEG (sEEG) and intracranial EEG (iEEG) based on the signal acquisition location. While sEEG is readily available and non-invasive, it is more prone to artifacts caused by electrode shifts, muscle movements, volume conduction effects, etc\cite{parvizi2018human}. In contrast, iEEG offers superior signal-to-noise ratio and sensitivity, as it directly targets specific brain areas and samples from deep brain regions inaccessible to scalp electrodes, at the cost of requiring brain surgery\cite{karoly2021cycles}.

Accurate decoding of epileptic EEG signals is crucial for aiding medical diagnosis, assisting neurologists in treatment, and reducing the risk of seizures\cite{saab2020weak}. Although EEG-based seizure detection has achieved significant progresses, these advances have imposed a considerable burden on physicians, requiring them to visually scrutinize up to several days of EEG signals to identify abnormal electrical discharges\cite{ney2013continuous}. Modern deep learning approaches highly rely on large hand-labeled datasets, which has been a significant bottleneck in healthcare applications of deep learning \cite{Esteva2019guide}. Detecting epilepsy for a new dataset frequently demands labor-intensive annotation efforts specific to that dataset. Numerous medical machine learning models have demonstrated inflexibility in adapting to shifting patient populations, evolving disease manifestations, variations in acquisition hardware, and so on, rendering them less effective in practice. Furthermore, ensuring the model to generalize well over diverse patient populations with varying characteristics poses a significant challenge, often necessitating model retraining \cite{saab2020weak}. Such time-consuming and expertise-dependent process underscores the need for automated algorithms capable of reliable and transferable EEG-based epilepsy detection.

Previous studies involving traditional feature-based machine learning algorithms\cite{acharya2013automated,faust2015wavelet,acharya2015application} and deep learning algorithms\cite{acharya2018deep,hussein2019optimized,sun2022continuous} have been widely explored for automatic epilepsy detection. Most works in the literature focused on within-patient seizure classification\cite{acharya2018deep,hussein2019optimized,sun2022continuous}, in which the training and test sets typically originate from the same group of patients. However, such within-patient analysis could not be directly applied to unseen patients due to significant non-stationarity and individual differences of EEG signals\cite{wang2023tasa}. Consequently, researchers have started investigating cross-patient seizure detection, utilizing auxiliary patient's labeled data to relieve the reliance on target patient's labeled data\cite{wang2023tasa,rukhsar2023lightweight,zhang2024cross}. However, both within-patient and cross-patient seizure detection require an adequate amount of labeled data from the human patients for effective model training, which may not always be available.

To solve the above challenge, we propose to harness existing data from other species and build a cross-species and cross-modality transfer learning framework for epilepsy seizure detection. A few prior works delineated below have explored the correlations of epilepsy of humans and other species, through the lens of biological mechanisms.

Several biological models have been shown to possess cross-species applicability\cite{chen2023lipid}. Many neurological and psychiatric disorders impact both humans and animals. Advances in diagnostics and therapeutics in human neurology and psychiatry are often translatable to veterinary patients, and vice versa. For instance, photosensitive generalized seizures in baboons provide insights into similar seizures in humans\cite{devinsky2018cross}. Lipid accumulation in the brain is a significant pathological feature of epilepsy in both humans and mice\cite{chen2023lipid}. Human ischemic stroke gene expression biomarkers can be obtained from rat brain samples\cite{wang2016obtaining}. Similar patterns and evolutions of epileptiform discharges are observed in both human and mouse brains\cite{chan2019role}.

\begin{figure*}[htpb]
\includegraphics[width=\linewidth,clip]{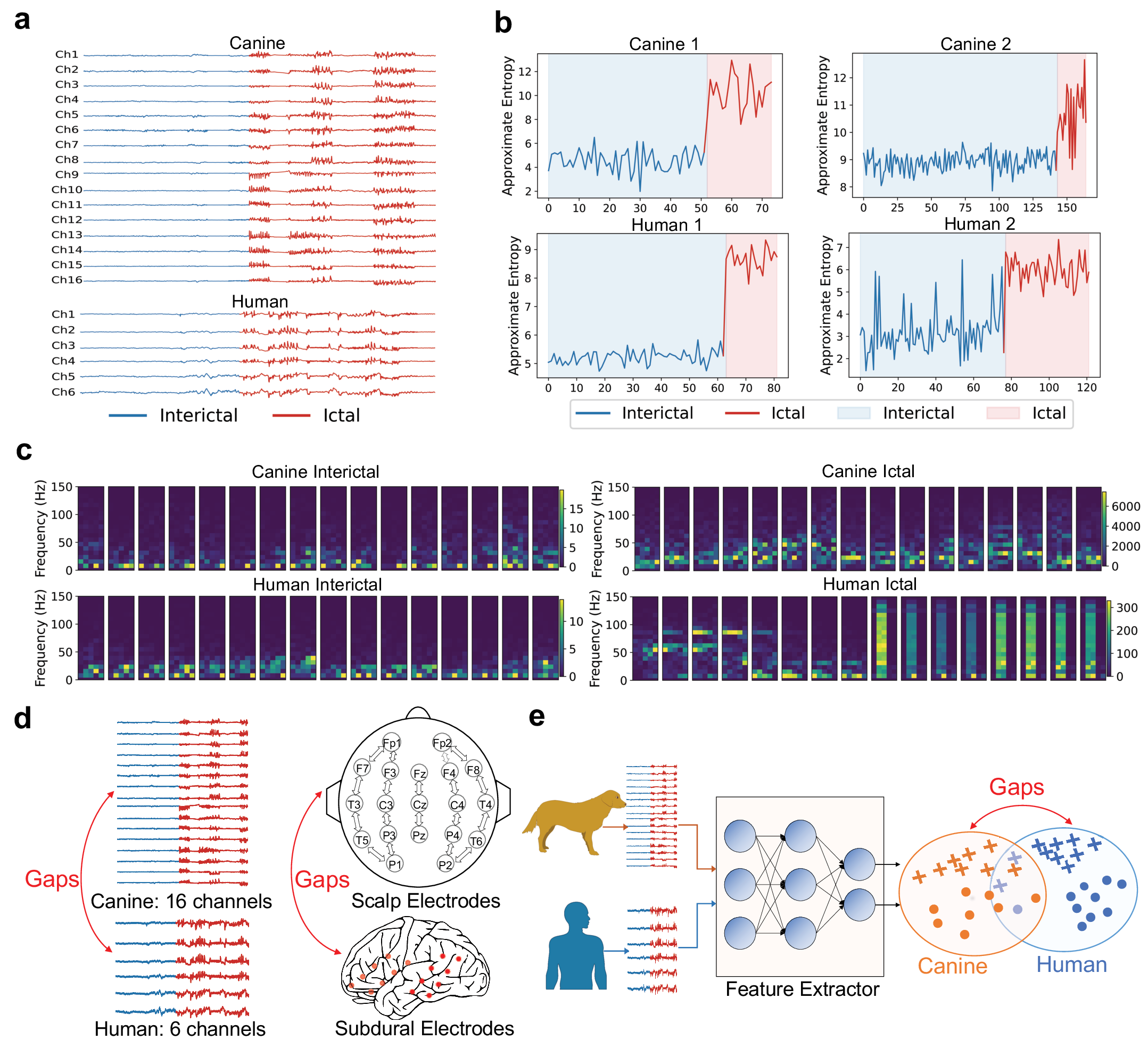}
\caption{\textbf{Evidences for cross-species and cross-modality feature transferability, along with gaps for successful knowledge transfer in algorithm design.} \textbf{a}, iEEG from both canines and humans exhibits large fluctuations during epileptic seizures, indicating the transferability of time-domain features across species. \textbf{b}, The approximate entropy of iEEG from both species increases significantly during seizures, indicating the transferability of entropy features across species. \textbf{c}, Power spectral density spectrograms, derived from consecutive Fourier transforms for both species, show an increase in the power across all channels during seizures, suggesting the transferability of frequency-domain features. \textbf{d}, Input space disparity across species are highlighted by the discrepancy in electrode configurations between species. Sixteen intracranial electrodes were used for canines iEEG data collection in the Kaggle dataset, whereas only six were used for humans iEEG data collection in the Freiburg dataset. Canine iEEG signals were captured using intracranial electrodes linked to implanted devices, whereas human sEEG signals in the NICU and CHSZ datasets were collected via scalp electrodes, demonstrating the configuration differences. \textbf{e}, Feature distribution gaps between canines and humans.} \label{fig:intro_cst}
\end{figure*}

Structural and functional elements of biological systems are shared among vertebrates\cite{devinsky2018cross}. Canine epilepsy shares many characteristics with human epilepsy\cite{berendt1999eeg}, including seizure clusters \cite{gregg2020circadian}, clinical presentation \cite{jeserevics2007eeg}, pharmacological features \cite{leppik2009canine}, periodic features\cite{karoly2021cycles, gregg2020circadian}, energy features, and temporal morphology \cite{pellegrino2004}, as illustrated in Figures~\ref{fig:intro_cst} a-c. For periodic features, Gregg \textit{et al.} \cite{gregg2020circadian} demonstrated that seizure periodicities and clusters are prevalent in canines, as in humans. For pharmacological features, Leppik \textit{et al.} \cite{leppik2009canine} observed that naturally occurring canine seizures closely mirror human epilepsy in pathophysiology, clinical progression, and response to conventional therapies. Additionally, canine studies on dosing and plasma drug concentrations for controlling seizures offer valuable insights for optimizing human treatment regimens. For brain wave characteristics, Pellegrino \textit{et al.} \cite{pellegrino2004} found that canine brain wave patterns closely resemble those of humans and apes, enabling cross-species comparisons to extrapolate healthy and seizure states. Furthermore, canine's brain size makes them compatible with iEEG devices designed for human use \cite{kremen2018integrating}. Previous works have shown that human drugs such as levetiracetam and gabapentin are of benefit to canines with refractory epilepsy\cite{chandler2006canine}, and naturally occurring canine epilepsy offers a potential biological model of human epilepsy\cite{holliday1970}. Adapted human and murine assays are effectively employed to examine the presence of neural antibodies in canines with idiopathic epilepsy \cite{hemmeter2023}. Similarities are found in diagnostic biomarkers between the human and canine species \cite{garcia2024}. Cross-species approaches to brain research can utilize prior knowledge from other species to enhance the understanding of diseases in the current species\cite{devinsky2018cross}.

\renewcommand{\dblfloatpagefraction}{.6}

We aim to utilize cross-species and cross-modality epileptic EEG signals to increase the amount of training data, investigate the common characteristics of epileptic signals across species and modalities, and facilitate knowledge transfer. We demonstrate that EEG data from one species/modality can enhance the seizure detection performance of another species/modality, suggesting the potential of integrating data from different species and modalities to improve EEG decoding performance. The proposed cross-species and cross-modality epilepsy seizure detection framework is shown in Figure~\ref{fig:method_cst}a. The deep neural network model is trained on canine iEEG signals and subsequently tested on human iEEG/sEEG signals, achieving a transfer from canines to humans. Likewise, the transfer from humans to canines is also achievable.

Importantly, disparities exist in the EEG signals across species that must be appropriately addressed to achieve optimal transfer performance, which is the core problem we aim to solve in this work. EEG signals of different species are collected from varying numbers of electrodes, using different sampling rates and diverse collection devices, as illustrated in Figure~\ref{fig:intro_cst}d. These factors lead to distinct signal characteristics, manifesting as differences in brain signals' temporal and spatial dimensions across different species and modalities. For instance, human iEEG signals could be sampled from various subdural electrode grids based on individual clinical considerations; the number of electrodes could go as many as 72, whereas canine iEEG signals may be acquired from an implanted device with 16 subdural electrodes\cite{brain2017kaggle}. In another example, the human iEEG signals were collected from three focal and three extra-focal electrodes\cite{Freiburg2012}, whereas sEEG signals were collected via scalp electrodes\cite{nicu2019, wang2023tasa}. These discrepancies complicate the cross-species transfer of feature heterogeneity. A na\"ive solution would be to discard the additional electrodes, resulting in the loss of valuable information. Even if the data are perfectly normalized in the input space, the extracted features may still exhibit significant distribution shifts, as illustrated in Figure~\ref{fig:intro_cst}e. Such problems have caused considerable difficulty in algorithm design for cross-species and cross-modality transfer.

In summary, the following technical challenges inherent in cross-species and cross-modality transfer must be systematically addressed:
\begin{enumerate}
\item Input heterogeneity across species and modalities: Differences in electrode configurations, sampling rates, and signal properties present significant obstacles to aligning the input spaces.
\item Distribution discrepancies across species, datasets, and subjects: Even after addressing the input heterogeneity, variations persist across input, feature, and output spaces. These differences pose additional challenges for effective transfer.
\item Limited labeled data in the target species, a common yet critical limitation in automatic seizure detection.
\end{enumerate}

To minimize cross-species and cross-modality discrepancies and enhance the transfer performance, we implement alignments simultaneously in the input, feature and output spaces, eliminating the gaps of channel heterogeneity, distribution shifts, and prediction inconsistency. The details are illustrated in Figure~ \ref{fig:method_cst} of Appendix A.

This study is important, because animal brain signals, such as canine EEG data, provide valuable complementary insights due to cross-species similarities in seizure patterns and underlying neural mechanisms. Utilizing cross-species and cross-modality data could enhance the decoding of EEG signals for species with limited available data. Additionally, merging EEG datasets from different species and modalities can substantially increase the quantity of the training set, providing adequate training data for building robust and general-purpose deep learning models or large-scale models. By exploring methods to mitigate data distribution discrepancies between species and modalities, we can also bridge inter-species and inter-modality gaps and offer novel insights for clinical endeavors.

\section{RESULTS AND ANALYSIS}\label{sec:res}
\subsection{Datasets introduction}
We aim to simulate realistic clinic settings where the target species in consideration provides little or no labeled data, and labeled data have to be obtained from auxiliary species to build accurate decoding algorithms. Such auxiliary datasets may have very distinct characteristics, including the number of channels, sampling rates, discrepancies in intracranial or scalp electrode placements, etc.

Four public epileptic seizure datasets were employed in this study: the Kaggle dataset from the Kaggle UPenn and Mayo Clinic's Seizure Detection Challenge\cite{brain2017kaggle}; the Freiburg dataset from the Epilepsy Center of the University Hospital of Freiburg, Germany\cite{Freiburg2012}; the CHSZ dataset from Wuhan Children's Hospital, Tongji Medical College, Huazhong University of Science and Technology, China\cite{wang2023tasa}; and, the NICU dataset from Helsinki University Hospital's neonatal intensive care unit\cite{nicu2019}. More details about the datasets can be found in Appendix B, and the data preprocessing procedures are outlined in Appendix C.

\subsection{Transfer tasks and scenarios}
We considered both canine-to-human transfer and human-to-canine transfer tasks:
\begin{enumerate}
\item Canine-to-human: Train the model using iEEG data from canines, and test on human iEEG/sEEG data.
\item Human-to-canine: Train the model using iEEG/sEEG data from humans, and test on canine iEEG data.
\end{enumerate}

Depending on whether the test species of interest has any labeled data, two cross-species transfer scenarios were experimented:
\begin{enumerate}
\item Unsupervised cross-species transfer: The training data are exclusively from another species, and data from the target species are utilized for testing. This is a typical unsupervised transfer learning scenario, with different species representing distinct domains.
\item Semi-supervised cross-species transfer: A small amount of labeled data from one subject in the target species is combined with all data from another species to train the model, with the remaining target data used for testing. This is a semi-supervised transfer learning scenario. For all datasets, the proportion of labeled trials per class for each subject increased from 5\% to 20\% with a step of 5\%.
\end{enumerate}

In the unsupervised cross-species transfer scenario, models were trained on auxiliary species' labeled data and tested on each subject of the target species, with the average classification scores across all target patients computed as the final result. In the semi-supervised cross-species transfer scenario, models were trained on auxiliary species' labeled data in combination with labeled data from a target subject, and tested on the remaining test data of the target subject, then averaged over all test subjects.

The Area Under the Receiver Operating Characteristic curve (AUC) metric was used as the performance measure. Compared with raw accuracy score, precision, or recall, the AUC is a better metric that evaluates a model's ability to discriminate between seizure and non-seizure clips, regardless of the specific classification threshold selected.

\subsection{Transfer learning approaches} \label{sec:baselines}
Euclidean Alignment (EA)\cite{He2020EA} was used as a standard preprocessing step on the two iEEG datasets to address the heterogeneities in electrode placements across subjects. EA standardizes iEEG data by aligning the mean covariance matrices of all iEEG trials for each subject individually, ensuring consistency across all subjects.

Domain adaptation\cite{Wilson2020} aims to enhance data distribution matching by integrating labeled source data with unlabeled target data. Five popular unsupervised domain adaptation approaches introduced in Appendix A were compared without the proposed ResizeNet introduced in Figure~\ref{fig:method_cst}b, and the number of channels for both species was unified by eliminating the mismatching channels.

Knowledge distillation\cite{Hinton2015KD} facilitates the transfer of knowledge from a more sophisticated model (teacher) to a simpler one (student). This process utilizes the Kullback-Leibler (KL) divergence to align the probability distributions of both models' outputs, ensuring similar predictive behaviors. Our approach incorporates channel-wise knowledge distillation to refine this alignment, depicted in more details in Methods and Figure~\ref{fig:method_cst}c. Six popular knowledge distillation approaches introduced in Appendix A were combined with the proposed ResizeNet to further improve the alignment performance.

We denote the proposed multi-space alignment (MSA) approach as ResizeNet+MSA, which simultaneously performs input, feature, and output space alignments to handle channel disparity and distribution shifts for cross-species and cross-modality transfer.

\subsection{Results}
The unsupervised cross-species and/or cross-modality transfer learning results are shown in Table~\ref{tab:cross_res}, the semi-supervised cross-species and within-modality transfer learning results are shown in Table~\ref{tab:semi_kaggleandfreiburg}, and the semi-supervised cross-species and cross-modality transfer learning results are shown in Table~\ref{tab:semi_chszandnicu}.

\noindent\textbf{Automatic epileptic seizure detection using cross-species and/or cross-modality deep learning achieves promising performance} For the two iEEG datasets (Kaggle and Freiburg), over 90\% AUC can be achieved with a minimum amount of target patient labeled data (as low as only 5\%), as shown in Tables~\ref{tab:semi_kaggleandfreiburg} and \ref{tab:semi_chszandnicu}. For the two sEEG datasets, transferring from humans to canines can also reach over 90\% AUC with 10\% target canine labeled data.

\noindent\textbf{Utilizing cross-species auxiliary labeled data is beneficial} Within-species models for seizure detection consistently under-performed models utilizing cross-species auxiliary labeled data. Within-species analysis highly relies on species-specific data and is impossible to perform in scenarios shown in Table~\ref{tab:cross_res}, since there are no labeled data from the target species. Even in scenarios shown in Tables~\ref{tab:semi_kaggleandfreiburg} and \ref{tab:semi_chszandnicu}, it is always beneficial to utilize auxiliary labeled data from another species, regardless of which specific algorithm was used. The results highlight the critical role of utilizing additional larger labeled training data in enhancing model generalizability and accuracy. Particularly, models trained exclusively within the same species failed to capture the broader phenotypic variations of epilepsy, which were effectively modeled by incorporating cross-species data, thus leveraging the generalized neurological patterns observed across different species.

\begin{figure*}[htpb]
\includegraphics[width=\linewidth,clip]{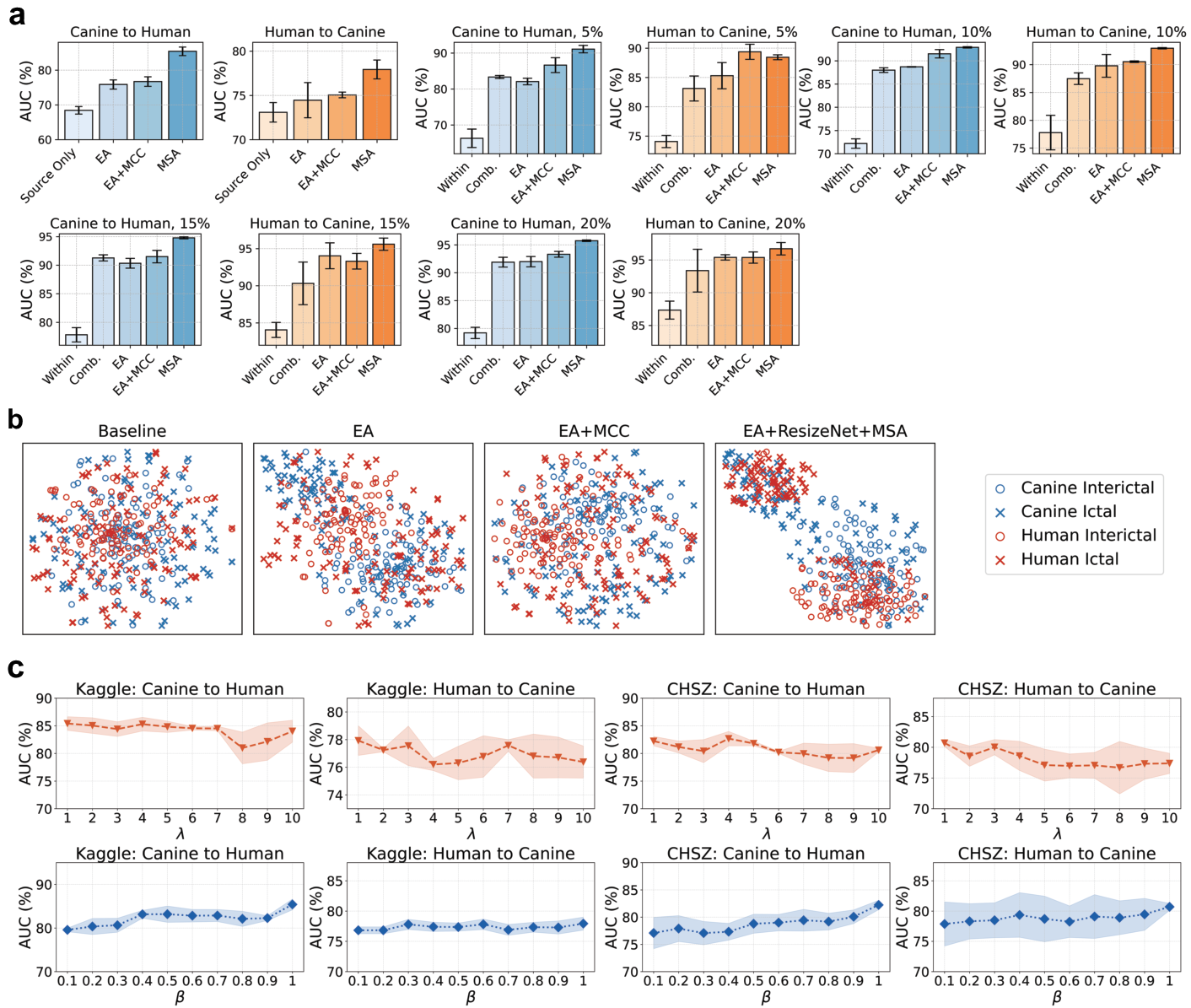}
\caption{\textbf{Ablation study, $t$-SNE feature visualization, and parameter sensitivity analysis}. \textbf{a}, Ablation study on the Kaggle iEEG dataset, including two tasks of canine-to-human and human-to-canine, and two scenarios of unsupervised cross-species transfer and semi-supervised cross-species transfer. \textbf{b}, Feature $t$-SNE visualizations on the Canine-to-Human cross-species transfer task of Kaggle, CHSZ and NICU datasets. \textbf{c}, Parameter sensitivity analysis of $\lambda$ and $\beta$ on the Kaggle iEEG dataset and CHSZ sEEG dataset using the proposed ResizeNet+MSA approach. A point denotes the average, and the shadow denotes the standard deviation. } \label{fig:ablation}
\end{figure*}

\noindent\textbf{Our MSA framework is much more effective than all standalone strategies} The MSA framework encompasses simultaneously input, feature, and output space alignments, demonstrating superior performance over all standalone strategies. As shown in the ablation study results in Figure~\ref{fig:ablation}a, performing alignments in more spaces generally resulted in higher AUCs. The basic baselines, ``Source Only" in the unsupervised transfer scenario and ``Within" or ``Comb." in the semi-supervised transfer scenario, consistently demonstrated the poorest performance across all tasks due to the absence of any alignment strategies. EA (input space alignment) and MCC (feature space alignment) were effective across most tasks. ResizeNet+KD and ResizeNet+MSA consistently achieved the second-best or best performance. While ResizeNet+KD focused on input space and output space alignments, ResizeNet+MSA extended this by performing alignment across the input, feature and output spaces, incorporating unsupervised domain adaptation beyond what ResizeNet+KD provides.

Figure~\ref{fig:ablation}b presents $t$-distributed stochastic neighbor embedding ($t$-SNE) \cite{VanderMaaten2008} visualizations of iEEG/sEEG features from the cross-species transfer task of Canine-to-Human on the Kaggle, CHSZ, and NICU datasets. In the two-dimensional reduced feature space, raw iEEG data from different species showed distinct distributions (Baseline). However, after alignment using EA, domain adaptation, or the proposed ResizeNet+MSA, the distributions became more similar, with significantly improved class separability (indicated by the merging of different colors for the same shape). Among these methods, ResizeNet+MSA achieved the best results, showing superior class separability (the circles were clearly distinguished from the crosses) and optimal alignment performance (blue and red points of the same shape were closely aligned).

We conducted hyperparameter sensitivity analysis to further investigate the impact of the two weights in Eq.~\eqref{eq:cstloss}, i.e., $\lambda$ for the knowledge distillation loss and $\beta$ for the domain adaptation loss. All experiments were repeated three times, and the results are shown in Figure~\ref{fig:ablation}c. The proposed ResizeNet with knowledge distillation loss was robust over a wide range of $\lambda$ and $\beta$ values.

\noindent\textbf{Cross-species and cross-modality feature heterogeneity can be accommodated by ResizeNet projection and knowledge distillation} The integration of ResizeNet and knowledge distillation addresses the cross-species and cross-modality feature heterogeneity effectively. ResizeNet's ability to project and transform features into a unified space, coupled with knowledge distillation's ability in refining and transferring essential information, proved crucial. As shown in Tables~\ref{tab:cross_res}-\ref{tab:semi_chszandnicu}, ResizeNet almost always outperformed Baseline approaches which brutally match the channels by truncation. ResizeNet ensured that the entire feature space pertinent to seizure detection was preserved and accurately interpreted across species, resulting in significant detection performance improvement.

Our proposed ResizeNet+MSA always achieved the best average performance in Tables~\ref{tab:cross_res} and \ref{tab:semi_kaggleandfreiburg}. Although in Table~\ref{tab:semi_chszandnicu} the average performance of ResizeNet+MSA was worse than the six ResizeNet based approaches in transferring from Canine (iEEG, Kaggle) to Human (sEEG, CHSZ), its average performance over the four transfer scenarios in Table~\ref{tab:semi_chszandnicu} was still the best among the 14 approaches (the average accuracy of ResizeNet+MSA was 87.45\%, and the average accuracy of the second best approach, ResiezeNet+CC, was 86.27\%). Moreover, ResizeNet+MSA was the only approach that consistently outperformed all Baselines, superior to all six other ResizeNet based approaches. These results demonstrated the effectiveness of MSA.

\begin{figure*}[htpb]
\includegraphics[width=\linewidth,clip]{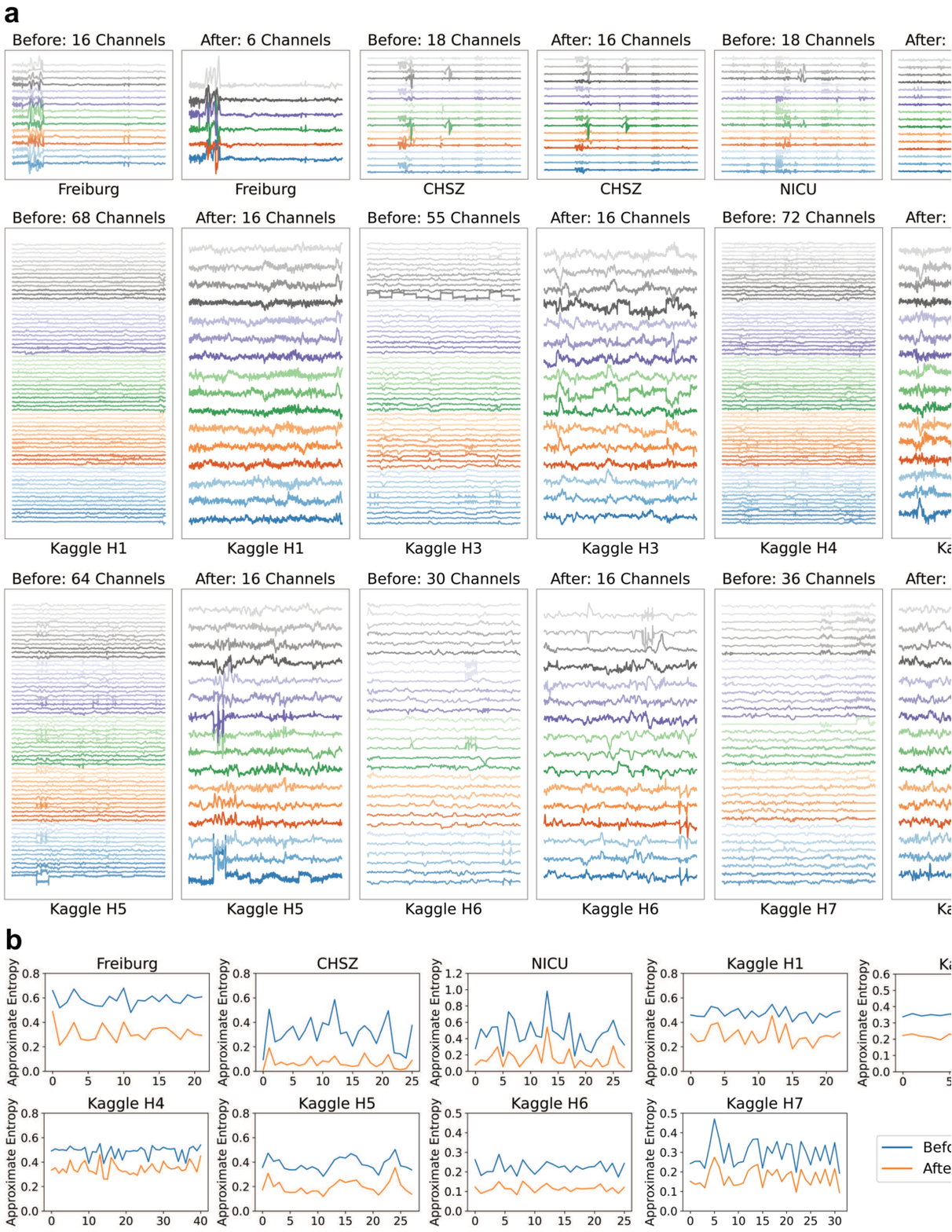}
\caption{\textbf{Effect of Input Space Alignment via the Proposed ResizeNet on the Four Datasets} (ResizeNet was not applied to the Kaggle H2 and H8 datasets, as their number of channels matches that of the canine dataset) \textbf{a}, The iEEG/sEEG signals before and after applying ResizeNet exhibit similar temporal structures, demonstrating its effectiveness. ResizeNet standardizes the number of channels across species, addressing input space discrepancies while preserving essential temporal structures and features. \textbf{b}, The mean approximate entropy of all channels before and after ResizeNet is shown for the four datasets, illustrating that ResizeNet retains critical entropy features of the iEEG/sEEG signals. \textbf{c}, Channel importance before and after applying ResizeNet, highlighting its ability to preserve significant spatial characteristics. Each column represents a different subject.} \label{fig:vis_isa}
\end{figure*}

\noindent\textbf{ResizeNet preserves essential classification information} To assess the impact of the proposed ResizeNet, Figure~\ref{fig:vis_isa}a shows the iEEG/sEEG trials in the time domain before and after input space alignment, Figure~\ref{fig:vis_isa}b illustrates their approximate entropy, and Figure~\ref{fig:vis_isa}c shows the spatial characteristics of channel importance before and after applying ResizeNet.

The importance of each EEG channel was evaluated using SHapley Additive exPlanations (SHAP) values \cite{scott2017shap}. SHAP, an additive explanation model inspired by cooperative game theory, quantifies each feature's (in this case, each channel's) contribution  to the model's predictions. Unlike traditional approaches for evaluating feature importance, which only identify significant features, SHAP also provides the magnitude and direction (positive or negative influence) of each feature's impact on individual predictions.

The channel importance calculation pipeline included three main steps. First, approximate entropy features were extracted from each channel, producing $C$ features for each signal, originally of dimensionality $(C,T)$. Next, a Random Forest classifier was trained on all available samples. Finally, SHAP values were calculated for each feature (channel), and the mean SHAP value across all samples was used to determine the importance of each channel. We utilized the TreeExplainer tool from \cite{lundberg2020shap} to generate the visualizations in Figure~\ref{fig:vis_isa}c. In the plot, each row corresponds to a feature representing a specific EEG channel. The horizontal axis shows the SHAP values, which quantify each channel's contribution to the model's predictions for individual samples. Each dot represents a sample, with its color reflecting the feature value (red for higher values and blue for lower values). Channels are ranked along the vertical axis in descending order of importance based on their relevance to the prediction outcomes.

The top row of Figure~\ref{fig:vis_isa}c presents SHAP-based channel importance for raw signals, and the bottom row shows the importance after applying ResizeNet for channel dimensionality alignment. For different subjects (different columns), the most important raw channel still exhibited high importance (among the top 3) after ResizeNet. For instance, channel 16 maintained high importance in the first column, channels 3 and 6 in the second column, channels 2 and 3 in the third column, and channels 15 and 16 in the fourth column. These patterns indicated that critical features were preserved post-resizing. Additionally, the SHAP value distribution remained largely consistent between the raw and resized signals. These results demonstrated that ResizeNet effectively retained the spatial characteristics of channel importance, ensuring robustness across various datasets and subjects.

In summary, Figure~\ref{fig:vis_isa} shows that iEEG/sEEG signals before and after ResizeNet exhibit similar temporal structures, entropy features, and spatial characteristics, demonstrating the effectiveness of ResizeNet in preserving the essential information while unifying diverse brain signals across species.

\noindent\textbf{EA is pivotal and a must-have for data normalization in all transfer learning tasks} EA proved to be essential in data normalization in cross-species and cross-modality transfers, addressing significant discrepancies in EEG data distributions. As shown in Figure~\ref{fig:ablation}a, the performance gain by using EA was tremendous. By aligning the data to a common scale, we effectively mitigated the discrepancies in the input space, which are crucial for subsequent machine learning. EA has been shown to be effective in various cross-subject transfer tasks\cite{drwuMITLBCI2022,Junqueira2024}, and this work shows that EA is also beneficial to cross-species and cross-modality transfers.

\noindent\textbf{Domain adaptation helps feature distribution alignment in cross-species  and cross-modality transfers} Domain adaptation aligns the feature distributions between species and modalities, significantly improving the transferability of seizure detection models, as shown in Tables~\ref{tab:cross_res}-\ref{tab:semi_chszandnicu}. By reducing the probability distribution gap, common and transferable features may be easily revealed.

\noindent\textbf{Our transfer learning framework significantly reduces expert labeling efforts} Our framework reduces the need for extensive expert labeling, which is a significant bottleneck in traditional EEG analysis for epilepsy seizure detection. Note that expert labeling for target patients is still of critical importance regardless of whether there exists auxiliary labeled data from other species and/or modalities: ResizeNet+MSA works better with a small portion of labeled data from the target subject than without any, e.g., results from semi-supervised cross-species transfer in Tables~\ref{tab:semi_kaggleandfreiburg}-\ref{tab:semi_chszandnicu} are much better than those in unsupervised cross-species transfer in Table~\ref{tab:cross_res}. Our approach cannot completely eliminate the need for expert labeling, but greatly reduces the amount of labeled target data to reach a specific accuracy. For example, as shown in Tables~\ref{tab:semi_kaggleandfreiburg}-\ref{tab:semi_chszandnicu}, cross-species transfer using ResizeNet+MSA with 5\% target labeled data generally outperformed within-species analysis of over 20\% labeled data.

\begin{table*}[htbp] \centering \setlength{\tabcolsep}{0.5mm} \footnotesize
\caption{Average unsupervised cross-species and/or cross-modality transfer AUCs (\%) on the four datasets. The best average AUC of each task is marked in bold, and the second best by an underline.} \label{tab:cross_res}
\begin{threeparttable}[b]
\begin{tabular*}{\textwidth}{@{\extracolsep\fill}c|c|ccccc|ccccc}
\toprule
\multicolumn{2}{@{}c|@{}}{\multirow{5}{*}{Approach}}& \multicolumn{5}{c|}{Canine-to-Human\tnote{1}} & \multicolumn{5}{c}{Human-to-Canine\tnote{2}} \\\cmidrule{3-7}\cmidrule{8-12}%
\multicolumn{2}{@{}c|@{}}{}& Kaggle & CHSZ & NICU & Freiburg & \multirow{4}{*}{Avg.} & Kaggle & CHSZ & NICU & Freiburg & \multirow{4}{*}{Avg.} \\
\multicolumn{2}{c|}{}&(iEEG & (iEEG &(iEEG & (iEEG & &(iEEG &(sEEG & (sEEG & (iEEG & \\
\multicolumn{2}{@{}c|@{}}{}&-to- & -to- &-to- & -to- & &-to- &-to- & -to- & -to- & \\
\multicolumn{2}{@{}c|@{}}{}&iEEG) & sEEG) &sEEG) & iEEG) & &iEEG) &iEEG) & iEEG) & iEEG) & \\
\midrule
\multirow{6}{*}{Baseline\tnote{3}}
& Source Only\tnote{5} & 68.43$_{\pm1.0}$ & 75.00$_{\pm2.7}$ & 62.87$_{\pm0.9}$ & 60.97$_{\pm0.5}$ & 66.82 & 73.09$_{\pm1.1}$ & 71.15$_{\pm2.5}$ & 53.26$_{\pm4.9}$ & 75.72$_{\pm1.8}$ & 68.31\\
& DAN & 75.91$_{\pm0.4}$ & 78.46$_{\pm0.5}$ & 68.76$_{\pm0.1}$ & 60.91$_{\pm0.7}$ & 71.01 & 73.69$_{\pm0.4}$ & 74.77$_{\pm1.5}$ & 66.56$_{\pm4.2}$ & 75.84$_{\pm0.5}$ & 72.72\\
& JAN & 75.47$_{\pm1.1}$ & 79.53$_{\pm1.6}$ & 67.99$_{\pm1.6}$ & 60.95$_{\pm1.1}$ & 70.99 & 73.66$_{\pm0.4}$ & 76.44$_{\pm1.3}$ & 60.73$_{\pm2.8}$ & 74.76$_{\pm1.0}$  & 71.40\\
& SHOT & 76.82$_{\pm0.2}$ & 68.16$_{\pm2.1}$ & \underline{69.72}$_{\pm0.5}$ & 59.70$_{\pm1.0}$ & 68.60 & 73.93$_{\pm0.6}$ & 58.05$_{\pm2.1}$ & 51.75$_{\pm0.5}$ & 66.64$_{\pm2.9}$ & 62.59\\
& DSAN & \underline{77.79}$_{\pm1.9}$ & 67.48$_{\pm5.5}$ & 67.03$_{\pm0.8}$ & 61.16$_{\pm0.7}$ & 68.37 & 73.71$_{\pm1.2}$ & 78.14$_{\pm1.1}$ & 59.07$_{\pm5.1}$ & 75.00$_{\pm0.6}$ & 71.48\\
& MCC & 76.71$_{\pm1.3}$ & 71.91$_{\pm2.0}$ & 69.54$_{\pm0.9}$ & 61.72$_{\pm0.9}$ & 69.97 &  75.05$_{\pm0.3}$ & 56.97$_{\pm1.6}$ & 53.64$_{\pm2.8}$ & 72.08$_{\pm2.0}$ & 64.44\\
\midrule
\multirow{7}{*}{ResizeNet+\tnote{4}}
& AT & 70.75$_{\pm1.9}$ & 79.31$_{\pm3.2}$ & 68.71$_{\pm2.2}$ & 61.00$_{\pm2.1}$ & 69.94 &  75.44$_{\pm0.2}$ & 77.27$_{\pm3.7}$ & \underline{66.68}$_{\pm4.5}$ & 74.04$_{\pm1.2}$ & \underline{73.36}\\
& NST & 73.98$_{\pm4.9}$ & 74.74$_{\pm1.5}$ & 66.22$_{\pm1.4}$ & \textbf{63.94}$_{\pm0.7}$ & 69.72 &  74.78$_{\pm1.7}$ & \underline{79.36}$_{\pm1.1}$ & 61.82$_{\pm6.6}$ & 75.75$_{\pm1.2}$ & 72.93\\
& SP & 74.05$_{\pm5.2}$ & 78.14$_{\pm4.0}$ & 68.51$_{\pm2.6}$ & \underline{63.59}$_{\pm1.1}$ & \underline{71.07} &  75.08$_{\pm0.5}$ & 78.07$_{\pm3.3}$ & 60.30$_{\pm1.6}$ & 73.93$_{\pm0.3}$ & 71.85\\
& RKD & 77.78$_{\pm1.0}$ & 78.70$_{\pm2.9}$ & 65.93$_{\pm1.5}$ & 61.39$_{\pm0.8}$ & 70.96 &  75.31$_{\pm0.4}$ & 76.17$_{\pm3.5}$ & 59.53$_{\pm2.7}$ & 75.71$_{\pm1.9}$ & 71.68\\
& PKT & 71.30$_{\pm3.5}$ & \underline{79.94}$_{\pm2.1}$ & 68.51$_{\pm1.6}$ & 63.49$_{\pm1.2}$ & 70.81 &  \underline{75.47}$_{\pm0.8}$ & 77.80$_{\pm3.1}$ & 61.03$_{\pm1.7}$ & \underline{75.85}$_{\pm2.1}$ & 72.54\\
& CC & 71.67$_{\pm0.6}$ & 79.15$_{\pm0.7}$ & 68.98$_{\pm1.9}$ & 57.20$_{\pm2.8}$ & 69.25 & 74.50$_{\pm1.6}$ & 76.09$_{\pm0.9}$ & 65.88$_{\pm2.2}$ & 75.15$_{\pm1.0}$ & 72.91\\
\cmidrule(r){2-12}
& MSA (ours) & \textbf{85.41}$_{\pm1.2}$ & \textbf{82.23}$_{\pm0.8}$ & \textbf{72.42}$_{\pm1.2}$ & 62.05$_{\pm0.6}$ & \textbf{75.53} & \textbf{77.93}$_{\pm1.0}$ & \textbf{80.70}$_{\pm0.6}$ & \textbf{72.10}$_{\pm1.6}$ & \textbf{76.83}$_{\pm1.6}$ & \textbf{76.89}\\
\bottomrule
\end{tabular*}
\begin{tablenotes}
\item[1]{Canine-to-Human: Train on iEEG data from four canines in the Kaggle dataset, and test on a human iEEG/sEEG dataset.}
\item[2]{Human-to-Canine: Train on a human iEEG/sEEG dataset, and test on iEEG data from four canines in the Kaggle dataset.}
\item[3]{Baseline: Without using the proposed ResizeNet, the number of channels for both species is unified by eliminating the mismatching ones.}
\item[4]{ResizeNet+: Utilize the proposed ResizeNet projection to unify the number of channels for different species and/or modalities.}
\item[5]{Source Only: Train the model on all labeled data from the other species without employing any alignment strategy.}
\end{tablenotes}
\end{threeparttable}
\end{table*}

\begin{table*}[htpb] \centering \setlength{\tabcolsep}{0.5mm} \footnotesize
\caption{Average semi-supervised cross-species and within-modality transfer AUCs (\%) on Kaggle and Freiburg iEEG datasets, with increasing amount of labeled data from the target species. The best average AUC of each task is marked in bold, and the second best by an underline.} \label{tab:semi_kaggleandfreiburg}
\begin{threeparttable}[b]
\begin{tabular*}{\textwidth}{@{\extracolsep\fill}c|c|ccccc|ccccc}
\toprule
\multicolumn{12}{@{}c@{}}{Cross-Species Transfer Between Canine (Kaggle) and Human (Kaggle)} \\
\midrule
\multicolumn{2}{@{}c|@{}}{\multirow{2}{*}{Approach}}& \multicolumn{5}{c|}{Canine (iEEG, Kaggle) to Human (iEEG, Kaggle)} & \multicolumn{5}{c}{Human (iEEG, Kaggle) to Canine (iEEG, Kaggle)} \\ \cmidrule{3-12}
\multicolumn{2}{c|}{} & 5\% & 10\% & 15\% & 20\% & Avg. & 5\% & 10\% & 15\% & 20\% & Avg.\\
\midrule
\multirow{7}{*}{Baseline}
& Within\tnote{1} & 66.28$_{\pm2.5}$ & 72.18$_{\pm1.0}$ & 77.81$_{\pm1.2}$ & 79.20$_{\pm1.0}$ & 73.87 & 74.08$_{\pm1.0}$ & 77.78$_{\pm3.1}$ & 84.05$_{\pm1.0}$ & 87.34$_{\pm1.3}$ & 80.81 \\
& Comb.\tnote{2} & 83.34$_{\pm0.4}$ & 87.99$_{\pm0.5}$ & 91.27$_{\pm0.5}$ & 91.90$_{\pm0.8}$ & 88.63 & 83.12$_{\pm2.1}$ & 87.46$_{\pm1.0}$ & 90.32$_{\pm2.8}$ & 93.37$_{\pm3.2}$ & 88.57 \\
& DAN & 81.96$_{\pm0.6}$ & 88.01$_{\pm0.6}$ & 87.32$_{\pm0.9}$ & 89.90$_{\pm1.1}$ & 86.80 & 84.29$_{\pm1.6}$ & 85.97$_{\pm2.6}$ & 91.64$_{\pm0.9}$ & 94.69$_{\pm1.1}$ & 89.15 \\
& JAN & 81.34$_{\pm1.1}$ & 86.79$_{\pm1.0}$ & 88.52$_{\pm0.2}$ & 89.69$_{\pm0.5}$ & 86.59 & 83.43$_{\pm0.8}$ & 87.37$_{\pm2.6}$ & 92.30$_{\pm1.5}$ & 93.85$_{\pm1.7}$ & 89.24 \\
& SHOT & 85.41$_{\pm1.6}$ & 86.97$_{\pm1.4}$ & 89.13$_{\pm0.9}$ & 90.90$_{\pm0.6}$ & 88.10 & 82.50$_{\pm1.5}$ & 87.14$_{\pm1.0}$ & 90.70$_{\pm0.5}$ & 91.94$_{\pm0.3}$ & 88.07 \\
& DSAN & 81.76$_{\pm0.9}$ & 84.73$_{\pm0.6}$ & 86.66$_{\pm1.0}$ & 86.94$_{\pm0.8}$ & 85.02 & 81.58$_{\pm1.7}$ & 86.12$_{\pm2.0}$ & 91.61$_{\pm0.7}$ & 94.09$_{\pm1.6}$ & 88.35 \\
& MCC & 86.64$_{\pm2.0}$ & 91.52$_{\pm0.9}$ & 91.50$_{\pm1.0}$ & 93.31$_{\pm0.5}$ & 90.74 & \textbf{89.37}$_{\pm1.2}$ & 90.52$_{\pm0.1}$ & 93.30$_{\pm1.0}$ & 95.37$_{\pm0.8}$ & 92.14 \\
\midrule
\multirow{7}{*}{ResizeNet+}
& AT & 89.29$_{\pm0.6}$ & 92.68$_{\pm0.5}$ & 94.09$_{\pm0.5}$ & \underline{95.88}$_{\pm0.3}$ & 92.99 &  86.16$_{\pm1.2}$ & 91.05$_{\pm0.9}$ & 94.22$_{\pm0.5}$ & \underline{97.34}$_{\pm0.7}$ & 92.19 \\
& NST & 88.72$_{\pm1.0}$ & 91.43$_{\pm0.7}$ & 94.13$_{\pm0.5}$ & 94.76$_{\pm0.5}$ & 92.26 &  85.24$_{\pm1.3}$ & 90.13$_{\pm2.1}$ & 92.17$_{\pm1.4}$ & 94.18$_{\pm1.1}$ & 90.43 \\
& SP & \underline{90.31}$_{\pm0.2}$ & \underline{93.08}$_{\pm0.7}$ & \underline{94.52}$_{\pm0.7}$ & 95.64$_{\pm0.0}$ & \underline{93.39} & 86.61$_{\pm1.1}$ & 91.22$_{\pm0.6}$ & 94.79$_{\pm1.5}$ & 96.84$_{\pm0.6}$ & 92.37 \\
& RKD & 88.55$_{\pm0.6}$ & 91.52$_{\pm0.6}$ & 92.49$_{\pm0.0}$ & 94.19$_{\pm0.5}$ & 91.69 & 85.97$_{\pm0.5}$ & 90.02$_{\pm1.8}$ & 92.22$_{\pm1.8}$ & 92.86$_{\pm1.7}$ & 90.27 \\
& PKT & 89.26$_{\pm0.7}$ & 92.63$_{\pm0.7}$ & 94.10$_{\pm0.4}$ & \textbf{96.12}$_{\pm0.6}$ & 93.03 &  85.81$_{\pm0.5}$ & 90.38$_{\pm1.7}$ & 95.12$_{\pm1.2}$ & \textbf{97.96}$_{\pm0.3}$ & 92.32 \\
& CC & 88.32$_{\pm2.2}$ & \textbf{93.16}$_{\pm0.4}$ & 94.08$_{\pm0.5}$ & 95.65$_{\pm0.5}$ & 92.80 & 87.19$_{\pm0.9}$ & \underline{92.09}$_{\pm0.4}$ & \underline{95.33}$_{\pm0.1}$ & 95.74$_{\pm0.0}$ & \underline{92.59} \\
\cmidrule(r){2-12}
& MSA (ours) & \textbf{91.10}$_{\pm1.0}$ & 92.90$_{\pm0.1}$ & \textbf{94.78}$_{\pm0.1}$ & 95.75$_{\pm0.1}$ & \textbf{93.63} & \underline{88.44}$_{\pm0.4}$ & \textbf{92.93}$_{\pm0.1}$ & \textbf{95.59}$_{\pm0.8}$ & 96.72$_{\pm0.9}$ & \textbf{93.42}\\
\bottomrule
\toprule
\multicolumn{12}{@{}c@{}}{Cross-Species Transfer Between Canine (Kaggle) and Human (Freiburg)} \\
\midrule
\multicolumn{2}{@{}c|@{}}{\multirow{2}{*}{Approach}}& \multicolumn{5}{c|}{Canine (iEEG, Kaggle) to Human (iEEG, Freiburg)} & \multicolumn{5}{c}{Human (iEEG, Freiburg) to Canine (iEEG, Kaggle)} \\ \cmidrule{3-12}
\multicolumn{2}{c|}{} & 5\% & 10\% & 15\% & 20\% & Avg. & 5\% & 10\% & 15\% & 20\% & Avg.\\
\midrule
\multirow{7}{*}{Baseline}
& Within & 62.13$_{\pm0.5}$ & 67.33$_{\pm0.2}$ & 72.00$_{\pm1.2}$ & 73.94$_{\pm0.6}$ & 68.85 &  74.08$_{\pm1.0}$ & 77.78$_{\pm3.1}$ & 84.05$_{\pm1.0}$ & 87.34$_{\pm1.3}$ & 80.81 \\
& Comb. & 83.86$_{\pm2.5}$ & 84.62$_{\pm0.3}$ & 85.84$_{\pm0.8}$ & 85.61$_{\pm0.6}$ & 84.98 &  87.03$_{\pm0.4}$ & 92.98$_{\pm0.4}$ & 95.47$_{\pm0.7}$ & 96.15$_{\pm0.3}$ & 92.91 \\
& DAN & \underline{89.81}$_{\pm1.3}$ & \underline{92.45}$_{\pm0.9}$ & \underline{93.16}$_{\pm1.2}$ & \underline{93.44}$_{\pm0.7}$ & \underline{92.22} & 85.26$_{\pm1.7}$ & 89.42$_{\pm1.7}$ & 93.68$_{\pm0.4}$ & 93.73$_{\pm0.5}$ & 90.52 \\
& JAN & 86.47$_{\pm1.3}$ & 89.65$_{\pm0.3}$ & 89.28$_{\pm0.1}$ & 90.81$_{\pm0.5}$ & 89.05 &  85.71$_{\pm0.3}$ & 90.12$_{\pm0.2}$ & 94.57$_{\pm0.4}$ & 94.79$_{\pm0.6}$ & 91.30 \\
& SHOT & 79.13$_{\pm0.4}$ & 82.49$_{\pm0.7}$ & 85.20$_{\pm0.6}$ & 84.60$_{\pm1.0}$ & 82.86 & 88.37$_{\pm1.1}$ & 91.55$_{\pm0.4}$ & 94.17$_{\pm0.6}$ & 94.76$_{\pm0.4}$ & 92.21\\
& DSAN & 84.87$_{\pm3.0}$ & 86.42$_{\pm1.5}$ & 86.22$_{\pm2.0}$ & 88.70$_{\pm1.4}$ & 86.55 &  80.52$_{\pm0.6}$ & 85.84$_{\pm1.1}$ & 89.95$_{\pm1.6}$ & 88.19$_{\pm2.8}$ & 86.13 \\
& MCC & 85.69$_{\pm0.8}$ & 86.59$_{\pm1.0}$ & 86.96$_{\pm0.5}$ & 87.79$_{\pm0.8}$ & 86.76 & 90.69$_{\pm0.6}$ & \underline{95.93}$_{\pm0.4}$ & 96.50$_{\pm0.3}$ & 97.29$_{\pm0.7}$ & \underline{95.10} \\
\midrule
\multirow{7}{*}{ResizeNet+}
& AT & 86.35$_{\pm1.1}$ & 89.07$_{\pm0.7}$ & 89.07$_{\pm0.5}$ & 89.38$_{\pm0.2}$ & 88.47 & 89.29$_{\pm4.3}$ & 94.58$_{\pm0.4}$ & 96.50$_{\pm0.9}$ & 96.91$_{\pm0.8}$ & 94.32\\
& NST & 84.55$_{\pm1.7}$ & 86.38$_{\pm1.1}$ & 87.32$_{\pm2.1}$ & 87.95$_{\pm2.3}$ & 86.55 &  88.17$_{\pm4.1}$ & 88.88$_{\pm3.3}$ & 93.74$_{\pm1.9}$ & 93.78$_{\pm2.7}$ & 91.14\\
& SP & 86.40$_{\pm1.6}$ & 89.73$_{\pm0.6}$ & 89.25$_{\pm1.1}$ & 89.66$_{\pm0.4}$ & 88.76 &  90.97$_{\pm3.0}$ & 92.47$_{\pm0.8}$ & 96.23$_{\pm1.1}$ & \underline{97.53}$_{\pm0.1}$ & 94.30\\
& RKD & 82.84$_{\pm1.8}$ & 88.06$_{\pm1.0}$ & 87.68$_{\pm2.2}$ & 88.99$_{\pm2.1}$ & 86.89&  88.84$_{\pm1.7}$ & 93.48$_{\pm2.3}$ & \underline{96.90}$_{\pm0.2}$ & 97.21$_{\pm0.4}$ & 94.11\\
& PKT & 86.47$_{\pm1.2}$ & 89.65$_{\pm0.3}$ & 89.15$_{\pm0.0}$ & 90.77$_{\pm0.5}$ & 89.01 &  \underline{91.59}$_{\pm2.3}$ & 94.37$_{\pm1.1}$ & 96.84$_{\pm1.2}$ & 97.32$_{\pm0.2}$ & 95.03\\
& CC & 86.12$_{\pm0.6}$ & 89.32$_{\pm0.6}$ & 89.57$_{\pm0.2}$ & 90.43$_{\pm0.7}$ & 88.86 &  89.53$_{\pm3.9}$ & 92.95$_{\pm2.0}$ & 96.33$_{\pm0.2}$ & 96.94$_{\pm0.6}$ & 93.94\\
\cmidrule(r){2-12}
& MSA (ours) & \textbf{91.06}$_{\pm0.9}$ & \textbf{93.90}$_{\pm1.2}$ & \textbf{94.78}$_{\pm1.0}$ &  \textbf{94.79}$_{\pm0.8}$ & \textbf{93.63}& \textbf{92.85}$_{\pm0.5}$ & \textbf{96.92}$_{\pm0.9}$ & \textbf{97.04}$_{\pm0.6}$ & \textbf{97.92}$_{\pm0.1}$ & \textbf{96.18}\\
\bottomrule
\end{tabular*}
\begin{tablenotes}
\item[1]{Within: This baseline trains the model solely on the $l$\% labeled iEEG data of the current species, without utilizing any labeled iEEG data from other species.}
\item[2]{Comb.: Train the model on the combination of $l$\% labeled iEEG data from the current species and all labeled iEEG data from the other species, without employing any alignment strategies.}
\end{tablenotes}
\end{threeparttable}
\end{table*}

\begin{table*}[htpb] \centering \setlength{\tabcolsep}{0.5mm}
\footnotesize
\caption{Average semi-supervised cross-species and cross-modality transfer AUCs (\%) on CHSZ and NICU sEEG datasets. The best average AUC of each task is marked in bold, and the second best by an underline.} \label{tab:semi_chszandnicu}
\begin{tabular*}{\textwidth}{@{\extracolsep\fill}c|c|ccccc|ccccc}
\toprule
\multicolumn{12}{@{}c@{}}{Cross-Species and cross-modality Transfer Between Canine (Kaggle) and Human (CHSZ)} \\
\midrule
\multicolumn{2}{@{}c|@{}}{\multirow{2}{*}{Approach}}& \multicolumn{5}{c|}{Canine (iEEG, Kaggle) to Human (sEEG, CHSZ)} & \multicolumn{5}{c}{Human (sEEG, CHSZ) to Canine (iEEG, Kaggle)} \\ \cmidrule{3-12}
\multicolumn{2}{c|}{} & 5\% & 10\% & 15\% & 20\% & Avg. & 5\% & 10\% & 15\% & 20\% & Avg.\\
\midrule
\multirow{7}{*}{Baseline}
& Within & 62.02$_{\pm0.6}$ & 64.02$_{\pm0.2}$ & 65.65$_{\pm1.2}$ & 70.14$_{\pm1.6}$ & 65.46& 74.08$_{\pm1.0}$ & 77.78$_{\pm3.1}$ & 84.05$_{\pm1.0}$ & 87.34$_{\pm1.3}$ & 80.81\\
& Comb. & 78.97$_{\pm1.4}$ & 80.24$_{\pm0.9}$ & 80.09$_{\pm2.4}$ & 79.93$_{\pm2.3}$ & 79.81& 83.84$_{\pm0.7}$ & 89.64$_{\pm0.4}$ & 91.31$_{\pm0.5}$ & 93.60$_{\pm0.9}$ & 89.60\\
& DAN & 74.24$_{\pm0.8}$ & 74.28$_{\pm1.8}$ & 76.43$_{\pm1.0}$ & 75.36$_{\pm2.5}$ & 75.08&  84.05$_{\pm0.2}$ & 90.48$_{\pm0.3}$ & \underline{92.89}$_{\pm1.0}$ & \underline{94.28}$_{\pm0.9}$ & 90.43\\
& JAN & 70.10$_{\pm2.0}$ & 69.03$_{\pm1.0}$ & 71.23$_{\pm0.9}$ & 72.16$_{\pm1.2}$ & 70.63&  76.83$_{\pm0.0}$ & 82.32$_{\pm1.4}$ & 86.56$_{\pm3.4}$ & 89.53$_{\pm3.1}$ & 83.81\\
& SHOT & 72.96$_{\pm2.2}$ & 73.20$_{\pm0.8}$ & 73.64$_{\pm1.7}$ & 74.48$_{\pm1.5}$ & 73.57&  79.74$_{\pm1.8}$ & 86.95$_{\pm0.0}$ & 88.84$_{\pm0.8}$ & 89.47$_{\pm1.1}$ & 86.25\\
& DSAN & 68.34$_{\pm1.1}$ & 70.86$_{\pm1.2}$ & 70.12$_{\pm0.5}$ & 71.79$_{\pm1.9}$ & 70.28&  82.12$_{\pm1.0}$ & 86.72$_{\pm6.7}$ & 88.15$_{\pm1.1}$ & 88.86$_{\pm0.3}$ & 86.46\\
& MCC & 76.32$_{\pm1.5}$ & 78.44$_{\pm0.5}$ & 78.68$_{\pm1.0}$ & 78.60$_{\pm1.4}$ & 78.01&  83.54$_{\pm2.0}$ & 89.73$_{\pm1.2}$ & 92.36$_{\pm0.8}$ & 93.56$_{\pm0.8}$ & 89.80\\
\midrule
\multirow{7}{*}{ResizeNet+}
& AT & \underline{85.00}$_{\pm1.3}$ & 85.87$_{\pm0.9}$ & 87.83$_{\pm0.5}$ & 88.11$_{\pm0.7}$ & 86.70&  82.99$_{\pm1.5}$ & 85.70$_{\pm1.3}$ & 89.99$_{\pm1.3}$ & 89.84$_{\pm0.9}$ & 87.13\\
& NST & 83.71$_{\pm0.8}$ & 83.85$_{\pm1.7}$ & 86.64$_{\pm0.1}$ & 88.45$_{\pm1.0}$ & 85.66&  \underline{85.35}$_{\pm1.8}$ & \underline{91.19}$_{\pm0.6}$ & 92.39$_{\pm1.0}$ & 93.70$_{\pm1.5}$ & \underline{90.66}\\
& SP & 83.35$_{\pm2.2}$ & \textbf{86.51}$_{\pm0.3}$ & \underline{88.26}$_{\pm0.8}$ & \underline{88.69}$_{\pm0.5}$ & 86.70 & 82.31$_{\pm2.6}$ & 87.50$_{\pm1.5}$ & 90.30$_{\pm1.5}$ & 93.01$_{\pm1.1}$ & 88.28\\
& RKD & 84.46$_{\pm1.0}$ & 85.25$_{\pm2.0}$ & 86.19$_{\pm0.4}$ & 87.94$_{\pm0.4}$ & 85.96 & 84.27$_{\pm4.3}$ & 88.79$_{\pm1.2}$ & 86.93$_{\pm1.1}$ & 92.74$_{\pm1.2}$ & 88.18\\
& PKT & \textbf{86.40}$_{\pm0.4}$ & \underline{86.25}$_{\pm0.5}$ & 87.36$_{\pm0.8}$ & \textbf{89.19}$_{\pm0.5}$ & \textbf{87.30} & 83.29$_{\pm2.3}$ & 87.36$_{\pm1.6}$ & 89.80$_{\pm1.5}$ & 92.92$_{\pm1.2}$ & 88.34\\
& CC & 84.35$_{\pm1.9}$ & 85.64$_{\pm2.1}$ & \textbf{88.63}$_{\pm1.4}$ & 88.35$_{\pm0.7}$ & \underline{86.74} &  85.20$_{\pm2.2}$ & 89.57$_{\pm0.8}$ & 90.92$_{\pm1.0}$ & 92.60$_{\pm1.7}$ & 89.57\\
\cmidrule(r){2-12}
& MSA (ours) & 82.74$_{\pm1.9}$ & 84.14$_{\pm1.8}$ & 84.84$_{\pm0.6}$ & 86.02$_{\pm1.2}$ & 84.44 &  \textbf{86.96}$_{\pm1.5}$ & \textbf{91.22}$_{\pm1.9}$ & \textbf{94.93}$_{\pm2.3}$ & \textbf{94.79}$_{\pm1.9}$ & \textbf{91.98}\\
\bottomrule
\toprule
\multicolumn{12}{@{}c@{}}{Cross-Species and cross-modality Transfer Between Canine (Kaggle) and Human (NICU)} \\
\midrule
\multicolumn{2}{@{}c|@{}}{\multirow{2}{*}{Approach}}& \multicolumn{5}{c|}{Canine (iEEG, Kaggle) to Human (sEEG, NICU)} & \multicolumn{5}{c}{Human (sEEG, NICU) to Canine (iEEG, Kaggle)} \\ \cmidrule{3-12}
\multicolumn{2}{c|}{} & 5\% & 10\% & 15\% & 20\% & Avg. & 5\% & 10\% & 15\% & 20\% & Avg.\\
\midrule
\multirow{7}{*}{Baseline}
& Within & 62.13$_{\pm0.5}$ & 67.33$_{\pm0.2}$ & 72.00$_{\pm1.2}$ & 73.94$_{\pm0.6}$ & 68.85& 74.08$_{\pm1.0}$ & 77.78$_{\pm3.1}$ & 84.05$_{\pm1.0}$ & 87.34$_{\pm1.3}$ & 80.81\\
& Comb. & 71.01$_{\pm0.7}$ & 74.01$_{\pm0.3}$ & 76.51$_{\pm0.2}$ & 78.90$_{\pm0.7}$ & 75.11& 84.82$_{\pm0.7}$ & 85.88$_{\pm1.6}$ & 87.02$_{\pm2.8}$ & 92.08$_{\pm1.0}$ & 87.45\\
& DAN & 70.37$_{\pm0.3}$ & 73.45$_{\pm0.5}$ & 75.57$_{\pm0.4}$ & 77.82$_{\pm1.0}$ & 74.30&  \underline{88.94}$_{\pm1.4}$ & \underline{94.27}$_{\pm1.0}$ & 92.44$_{\pm1.8}$ & 94.52$_{\pm1.3}$ & 92.54\\
& JAN & 67.90$_{\pm1.1}$ & 71.23$_{\pm0.7}$ & 74.28$_{\pm0.3}$ & 75.26$_{\pm0.5}$ & 72.17&  84.05$_{\pm0.4}$ & 92.31$_{\pm0.1}$ & 93.19$_{\pm0.7}$ & 94.14$_{\pm1.0}$ & 90.92 \\
& SHOT & 65.18$_{\pm1.4}$ & 69.57$_{\pm1.2}$ & 73.06$_{\pm0.2}$ & 73.80$_{\pm0.3}$ & 70.40&  77.74$_{\pm1.0}$ & 86.17$_{\pm0.8}$ & 88.65$_{\pm1.7}$ & 89.80$_{\pm1.7}$ & 85.59 \\
& DSAN & 66.44$_{\pm0.7}$ & 70.63$_{\pm0.2}$ & 71.03$_{\pm1.0}$ & 72.76$_{\pm0.2}$ & 70.22&  75.26$_{\pm2.0}$ & 85.57$_{\pm1.4}$ & 87.36$_{\pm1.7}$ & 82.20$_{\pm3.3}$ & 82.60 \\
& MCC & 66.18$_{\pm0.4}$ & 71.33$_{\pm0.9}$ & 74.75$_{\pm1.2}$ & 77.01$_{\pm0.1}$ & 72.32&  83.55$_{\pm0.5}$ & 92.21$_{\pm0.8}$ & 94.44$_{\pm0.5}$ & \underline{95.67}$_{\pm0.9}$ & 91.47 \\
\midrule
\multirow{7}{*}{ResizeNet+}
& AT & 74.28$_{\pm2.6}$ & \underline{77.87}$_{\pm2.6}$ & 79.12$_{\pm2.5}$ & \underline{81.17}$_{\pm3.0}$ & \underline{78.11} &  85.11$_{\pm0.5}$ & 91.22$_{\pm0.8}$ & 91.75$_{\pm1.8}$ & 92.79$_{\pm1.6}$ & 90.22 \\
& NST & 70.61$_{\pm0.9}$ & 72.70$_{\pm1.6}$ & 75.60$_{\pm1.5}$ & 76.42$_{\pm1.7}$ & 73.83 &  \textbf{90.08}$_{\pm1.3}$ & \textbf{94.34}$_{\pm1.5}$ & \underline{94.92}$_{\pm1.2}$ & 94.74$_{\pm1.7}$ & \textbf{93.52}\\
& SP & 74.54$_{\pm0.6}$ & 76.90$_{\pm1.5}$ & 78.17$_{\pm1.9}$ & 80.77$_{\pm2.6}$ & 77.60 &  85.29$_{\pm0.8}$ & 89.57$_{\pm0.4}$ & 91.42$_{\pm1.8}$ & 93.04$_{\pm2.6}$ & 89.83 \\
& RKD & 72.13$_{\pm0.8}$ & 76.74$_{\pm0.7}$ & 76.18$_{\pm0.8}$ & 79.35$_{\pm1.2}$ & 76.10 &  84.26$_{\pm2.4}$ & 86.85$_{\pm0.4}$ & 90.67$_{\pm0.7}$ & 91.87$_{\pm1.4}$ & 88.41 \\
& PKT & \underline{75.21}$_{\pm0.8}$ & 77.46$_{\pm1.8}$ & 78.82$_{\pm1.5}$ & 80.57$_{\pm2.7}$ & 78.02 & 84.53$_{\pm0.1}$ & 91.95$_{\pm1.9}$ & 91.35$_{\pm0.1}$ & 92.23$_{\pm1.8}$ & 90.02 \\
& CC & 73.51$_{\pm3.1}$ & 74.24$_{\pm2.6}$ & \underline{79.66}$_{\pm0.7}$ & 80.26$_{\pm2.4}$ & 76.92 &  86.55$_{\pm1.0}$ & 90.84$_{\pm2.1}$ & \textbf{95.40}$_{\pm0.2}$ & 94.60$_{\pm0.5}$ & 91.85 \\
\cmidrule(r){2-12}
& MSA (ours) & \textbf{75.48}$_{\pm0.5}$ & \textbf{79.07}$_{\pm0.4}$ & \textbf{82.56}$_{\pm0.7}$ & \textbf{84.33}$_{\pm0.2}$ & \textbf{80.36} & 88.03$_{\pm1.3}$ & 93.87$_{\pm1.4}$ & 93.91$_{\pm3.0}$ & \textbf{96.21}$_{\pm1.1}$ & \underline{93.01}\\
\bottomrule
\end{tabular*}
\end{table*}

\section{DISCUSSION}\label{sec:dis}
This study introduces a novel cross-species and cross-modality framework for epileptic seizure detection using EEG data. Through MSA, we have developed a framework that not only addresses the fundamental challenges of EEG data variability across species and modalities, but also enhances the generalizability and accuracy of seizure detection.

\subsection{Broad implications}
To our knowledge, this is the first study that demonstrates the effectiveness of integrating heterogeneous data from different species and modalities to improve EEG-based seizure detection performance. However, the implications of this research extend beyond improving epilepsy diagnostics. The cross-species and cross-modality framework could revolutionize how neurological disorders are studied and treated, offering a model for developing diagnostic tools that are more broadly applicable. For clinical practice, this means diagnosis for populations where traditional approaches are incapable may be performed. This pilot study offers valuable insights into the challenges and opportunities associated with integrating multi-species and multi-modality data. It lays the foundation for future efforts to collect large-scale EEG datasets, enabling the training of large brain models.

Our approach may also be generalizable to different brain-computer interface paradigms\cite{drwuMITLBCI2022,drwuPIEEE2023}, and suggests the possibility to combine data from different species/modalities to increase the amount of training data for large EEG models. The performance of the proposed ResizeNet+MSA approach on the seizure prediction task is presented in Table~\ref{tab:aesp_result}, demonstrating its effectiveness in this application.

\subsection{Limitations and future research}
Despite the promising performance, several limitations must be addressed in future research.

The diversity of EEG acquisition protocols and hardware configurations across different clinical and research settings introduces variability that complicates data standardization and model training. Establishing universal protocols for EEG data collection and preprocessing could enhance the reproducibility of our findings and facilitate wider adoption of the technology.

The current study primarily focuses on iEEG/sEEG data from canines and humans. Expanding this research to include other species and modalities (e.g., magnetoencephalography) could provide deeper insights into the neurological underpinnings of seizures and other related disorders.

Finally, merging multiple auxiliary datasets is also worth exploring. In the proposed cross-species epilepsy detection framework, the training set consists of one dataset from another species (and a small amount of data from the target species in the semi-supervised cross-species and cross-modality transfer scenario). However, in practice, multiple datasets could be utilized to augment the training set. For instance, a model can be trained on a combination of several human datasets and subsequently tested on the canine dataset.

\section{METHODS}\label{sec:method}
Disparity in data characteristics poses a significant challenge in real-world machine learning applications, particularly in EEG-based seizure detection. In cross-species seizure detection, where different species represent distinct domains, the source and target species often differ significantly in aspects such as seizure subtype, location, duration, and signal collection device. These variations introduce discrepancies across input, feature, and output spaces, compounded by individual differences and configuration discrepancies arising from diverse EEG equipment and protocols. For instance, the canine iEEG data in Kaggle used 16 channels, whereas the human iEEG data used 16 to 72 channels with varying sampling rates. Accommodating these discrepancies is crucial for optimal cross-species transfer performance.

We introduce our MSA approach below.

\subsection{Input space normalization}

Inter-species variability leads to significant discrepancies in the marginal probability distributions of EEG signals. Euclidean Alignment (EA)\cite{He2020EA} has been demonstrated to be effective in EEG-based classification tasks like motor imagery. This unsupervised approach aligns EEG data from different patients within the same species and across species, making subsequent analyses easier and more robust.

We apply EA to each subject separately to mitigate individual differences, and then combine all aligned EEG trials from the source subjects into a single source domain.

For a subject with $N$ EEG trials $\{X^n\}_{n=1}^N$, EA first computes their mean covariance matrix:
\begin{align}
\bar{R}=\frac{1}{N} \sum_{n=1}^{N} X^{n}\left(X^{n}\right)^{\top},
\end{align}
and then performs the alignment by
\begin{align}
\tilde{X}^{n}=\bar{R}^{-1 / 2} X^{n}.
\end{align}
$\tilde{X}^{n}$ then replaces $X^n$ in all subsequent operations.

This process is repeated for each subject in the source and target species to normalize its mean covariance matrix to the identity matrix, making them more consistent.

\subsection{Input space alignment}

As illustrated in Figure~\ref{fig:method_cst}b, addressing the disparity in channel dimensionality is pivotal to effective cross-species transfer. Our proposed solution combines a Transformer Encoder with a linear layer in a neural network architecture, referred to as ResizeNet.

The Transformer Encoder captures the intricate temporal dependencies and spatial relationships in the EEG data, reflecting the inter-connections between distinct channel locations using a self-attention mechanism. Subsequently, a linear layer is employed to project EEG data from a higher dimensionality to a lower space, ensuring consistent input to the subsequent feature extractor.

Let the original EEG data be represented as $X \in \mathbb{R}^{N \times C \times T_s}$, where $N$ is the training batch size, $C$ the number of channels, and $T_s$ the number of time samples. The EEG data are first reshaped to $X \in \mathbb{R}^{N \times T_s \times C}$, and then input to the Transformer Encoder, which includes multi-head self-attention and a feedforward neural network. After this, a linear layer reduces the dimensionality for species with more channels. For instance, the human sEEG data in the CHSZ dataset have 18 channels, whereas the canine iEEG data in the Kaggle dataset have 16 channels; when transferring from the CHSZ dataset to the Kaggle dataset, ResizeNet reduces the human sEEG data to 16 channels to match the dimensionality of the canine iEEG data.

The final ResizeNet transformation is expressed as:
\begin{align}
\hat{X} = R(T(R(X)) \cdot W_L),\label{eq:resize}
\end{align}
where $R(\cdot)$ is the reshape function used for switching the last two dimensions of data, $T(\cdot)$ the Transformer Encoder with two layers and two heads, and $W_L \in \mathbb{R}^{C \times C_t}$ the learned parameter matrix, with $C_t$ being the target number of channels.

\subsection{Feature space alignment}

After ResizeNet alignment that uniforms the input signal dimensionality from both source and target species, domain adaptation can then be conducted to further align the feature distributions.

Domain adaptation leverages data from the labeled source domain $\mathcal{D}_s=\{(\mathbf{X}_s^i,y_s^i)\}_{i=1}^{N_s}$ and the unlabeled target domain $\mathcal{D}_t=\{(\mathbf{X}_t^i)\}_{i=1}^{N_t}$ to minimize their discrepancies, enabling the model trained on the source domain to generalize to the target domain. Traditional domain adaptation methods mainly consider the simpler homogeneous domain adaptation, i.e., the feature spaces of the source and target domains are identical ($\mathcal{X}_s=\mathcal{X}_t$), but their probability distributions are different ($P(X_s,y_s) \neq P(X_t,y_t)$).

The loss function of domain adaptation using MMD is:
\begin{align}
L_{\mathrm{DA}}&=\mathrm{MMD}^{2}\left(X_{s}, \hat{X}_{t}\right) \nonumber\\
&=\left\|\frac{1}{N_s} \sum_{i=1}^{N_s} \phi\left(\mathrm{x}_{i}^{s}\right)-\frac{1}{N_t} \sum_{j=1}^{N_t} \phi\left(\mathrm{\hat{x}}_{j}^{t}\right)\right\|_{\mathcal{H}}^{2},
\end{align}
where $\mathcal{H}$ is a reproducing kernel Hilbert space with a feature mapping $\phi$, and $\hat{X}_{t}$ is the ResizeNet transformation of ${X}_{t}$ defined in Eq.~\eqref{eq:resize}. Extensions of MMD, e.g., joint MMD\cite{Long2017JAN} and local MMD\cite{Zhu2020DSAN}, can also be utilized in domain adaptation regularization.

\subsection{Output space alignment}

We utilizes knowledge distillation to further align the output spaces. Knowledge distillation forces the student model to mimic the teacher model's behavior by imposing a stringent congruent constraint on their predictions, typically utilizing the Kullback–Leibler divergence.

As illustrated in Figure~\ref{fig:method_cst}c, the original EEG signals with $C$ channels are first reduced to $C_t$ channels by ResizeNet, or through simple channel selection (the first $C_t$ channels were used in this paper). In this way, we obtain two different signals with the same shape, one by ResizeNet and the other by channel selection. A neural network is then used to compute $z_r$, the logits from ResizeNet, and $z_s$, the logits from channel selection. Subsequently, knowledge distillation is performed on $z_r$ and $z_s$:
\begin{align}
L_{\mathrm{KD}}=\frac{1}{N} \sum_{i=1}^{N} \tau^{2} K L\left(\boldsymbol{p}_{r}, \boldsymbol{p}_{s}\right),
\end{align}\label{eq:kd}
where $\tau$ is a relaxation parameter (referred to as the temperature in\cite{Hinton2015KD}) to soften the output of the teacher network, and $p=\frac{\exp \left(z_{k} / \tau\right)}{\sum_{j} \exp \left(z_{j} / \tau\right)}$ is the softmax operation, with $z_i$ being the logit for the $i$-th class and $n$ the total number of classes.

\subsection{Multi-space alignment}

To minimize the gaps between species, we perform the above input-space, feature-space, and output-space alignments simultaneously. In the proposed cross-species transfer learning framework shown in Figure~ \ref{fig:method_cst}a, we first perform EA and ResizeNet to match the input dimensionality of the source and target species. Then, domain adaptation is performed to further reduce the discrepancies in the feature space. Finally, knowledge distillation is performed on the logits from ResizeNet and channel selection strategy.

The overall optimization objective is composed of three parts: the supervised cross-entropy loss, the domain adaptation loss, and the knowledge distillation loss:
\begin{equation}
L=L_{\mathrm{CE}} + \lambda \cdot L_{\mathrm{KD}} + \beta \cdot L_{\mathrm{DA}},\label{eq:cstloss}
\end{equation}
where $L_{\mathrm{CE}}$ is the classic cross-entropy loss for supervised learning, and $\lambda$ and $\beta$ are trade-off hyperparameters.

To summarize, EA normalizes the input data to a unified scale, ResizeNet reduces higher-dimensional data to a lower-dimensional representation, domain adaptation aligns feature distributions, and knowledge distillation ensures consistency in model predictions. Together, these alignments form a unified framework tailored to address the complexities of cross-species and cross-modality EEG analysis. By integrating innovations across the input, feature, and output spaces, the proposed approach achieves effective data normalization, robust feature representation, and consistent predictive performance. The MSA framework is illustrated in Figure~\ref{fig:method_cst}.

\subsection{Implementation details}

EEGNet\cite{Lawhern2018EEGNet}, a popular end-to-end convolutional neural network for EEG signal decoding, was utilized as the feature extractor, along with a fully connected layer as the classifier. In unsupervised cross-species transfer, all data from the target species were unlabeled for testing. In semi-supervised cross-species transfer, the first $l$\% labeled data of all subjects from the target species were combined and utilized during training, and the remaining $(100-l)$\% were used for testing. The trade-off parameters $\lambda$ and $\beta$ were both set to 1 in all experiments. To avoid temporal leakage pointed out in previous works\cite{kapoor2023leakage}, we partitioned the training and test data from the target species chronologically instead of randomly.

All experiments were repeated three times, and the average results are reported. All algorithms were implemented in PyTorch.

\clearpage

\begin{appendix}

\section{APPENDIX A: ALGORITHMS UNDER COMPARISON}\label{secA1}

The proposed MSA framework is illustrated in Figure~\ref{fig:method_cst}. ResizeNet (Figure~\ref{fig:method_cst}b) incorporates a Transformer encoder, a linear layer, and two reshape modules. MSA (Figure~\ref{fig:method_cst}a) integrates alignments across input, feature, and output spaces for cross-species adaptation. Specifically, ResizeNet addresses input space discrepancies, domain adaptation aligns cross-species data in the feature space (Figure~\ref{fig:method_cst}d), and knowledge distillation ensures output space alignment between species (Figure~\ref{fig:method_cst}c). Details of the training and test set splits for the three experiment scenarios (within-species, unsupervised cross-species transfer, and semi-supervised cross-species transfer) are illustrated in Figure~\ref{fig:cross_species_settings}.

\setcounter{figure}{0}
\renewcommand{\thefigure}{A\arabic{figure}}
\renewcommand{\dblfloatpagefraction}{.9}
\begin{figure*}[htpb]
\includegraphics[width=\linewidth,clip]{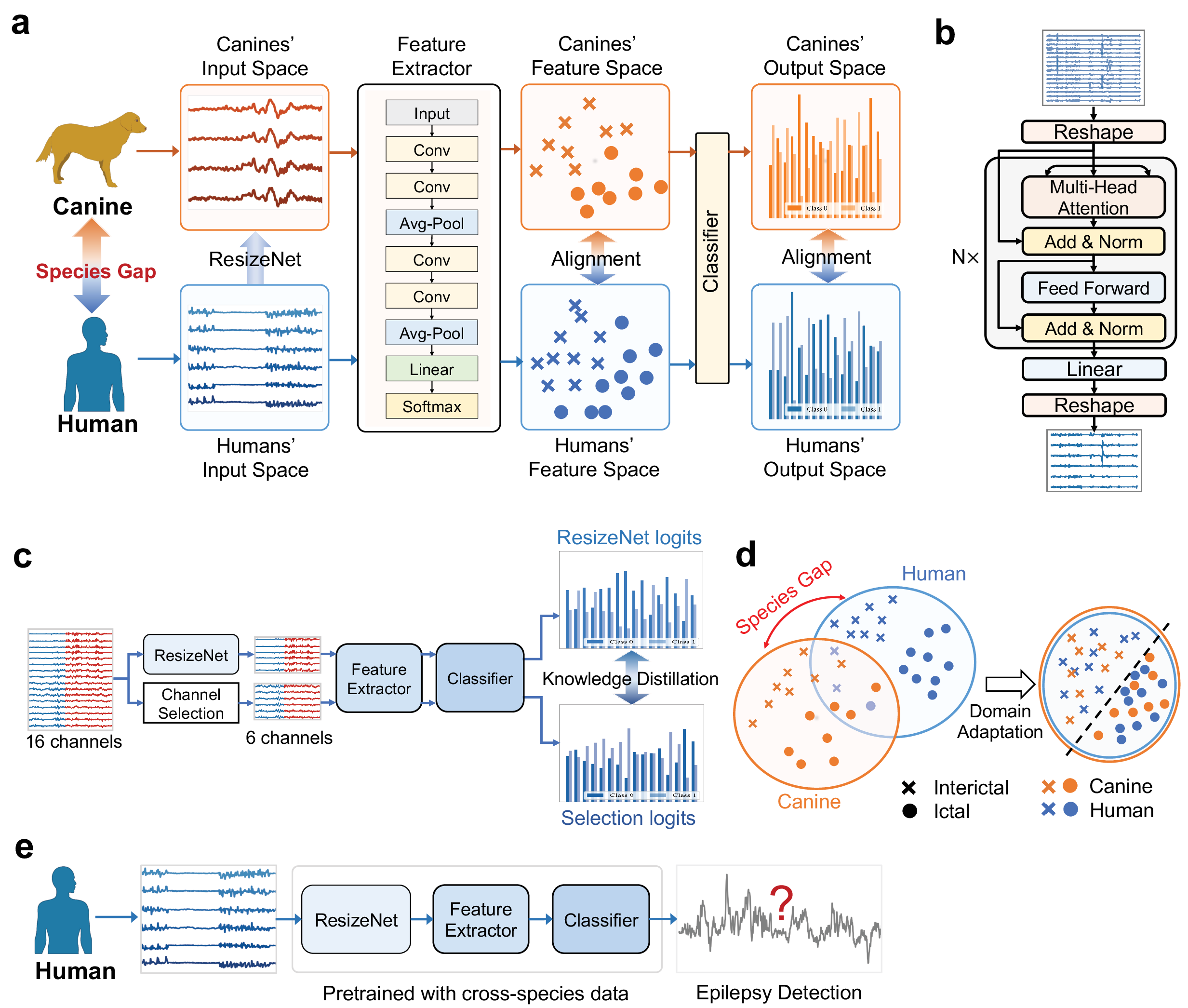}
\caption{\textbf{Overview of the proposed cross-species and cross-modality epilepsy seizure detection framework.} \textbf{a}, Training of the cross-species and cross-modality transfer network utilizes iEEG/sEEG data from both canines and humans. \textbf{b}, The proposed ResizeNet, which projects EEG signal of the species with higher dimensionality (collected with more EEG electrodes) to a lower dimensionality to match their feature spaces. \textbf{c}, The integration of knowledge distillation with ResizeNet for output-space alignment. \textbf{d}, Domain adaptation for distribution matching to achieve feature-space alignment. \textbf{e}, Illustration of the test phase on human iEEG/sEEG data.} \label{fig:method_cst}
\end{figure*}

\setcounter{figure}{1}
\renewcommand{\thefigure}{A\arabic{figure}}
\renewcommand{\dblfloatpagefraction}{.9}
\begin{figure*}[htpb]
\includegraphics[width=\linewidth,clip]{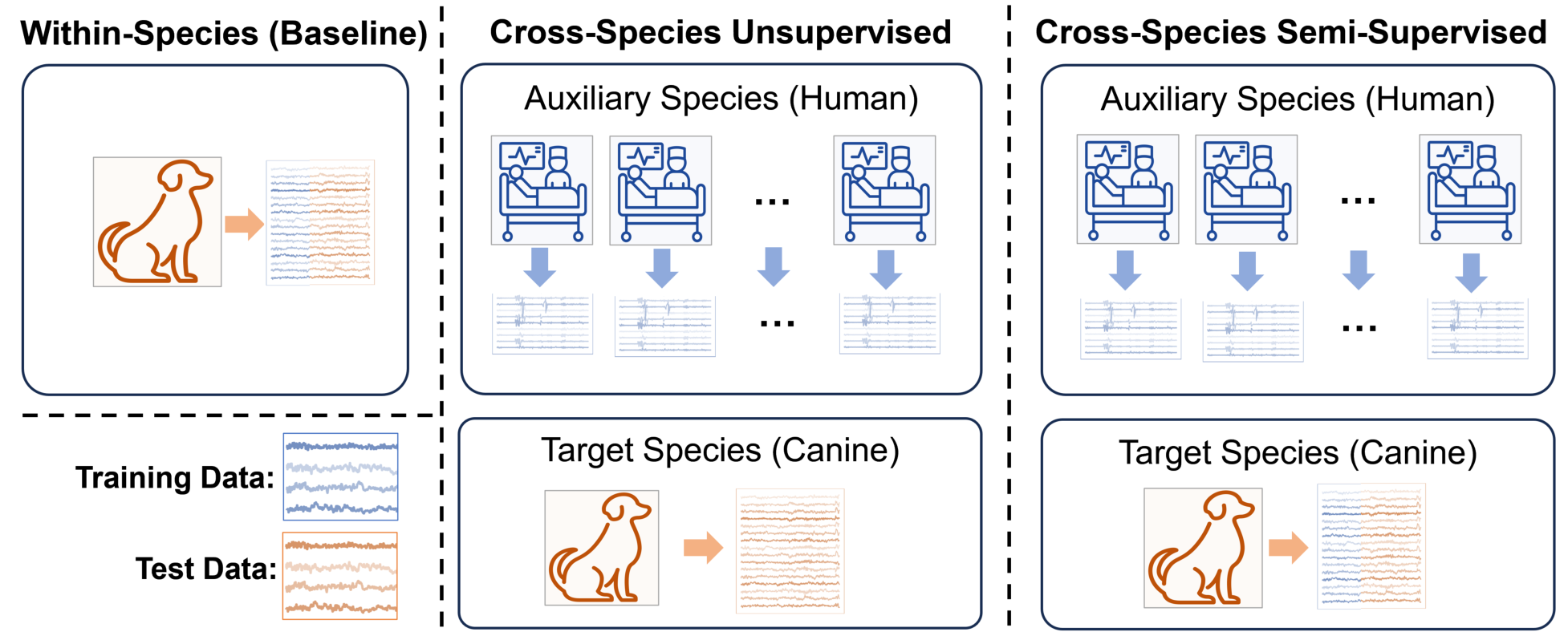}
\caption{Illustration of three experiment scenarios, taking the Human-to-Canine transfer task as an example.} \label{fig:cross_species_settings}
\end{figure*}

Five popular unsupervised domain adaptation approaches were compared:
\begin{enumerate}
\item Deep Adaptation Network (DAN)\cite{Long2015DAN}, which achieves feature alignment by minimizing the maximum mean discrepancies (MMD)\cite{Gretton2012} in the feature space.
\item Joint Adaptation Network (JAN)\cite{Long2017JAN} and Deep Subdomain Adaptation Network (DSAN)\cite{Zhu2020DSAN}, which construct transfer networks by aligning class-conditioned subdomain distributions based on local MMD in the feature space, utilizing either model prediction probabilities or pseudo-labels.
\item Minimum class confusion (MCC)\cite{Jin2020MCC}, which integrates a weighted prediction entropy with class correlation to reduce class confusion in the output space.
\item Source hypothesis transfer (SHOT)\cite{Liang2022SHOT}, which utilizes information maximization for conditional entropy minimization and label marginal entropy regularization in the output space.
\end{enumerate}

Six popular and diverse knowledge distillation approaches were combined with the proposed ResizeNet to further improve the alignment performance:
\begin{enumerate}
\item Attention Transfer (AT)\cite{Zagoruyko2017AT}, which utilizes an attention mechanism to enhance the student network's performance by mapping attention from the teacher's feature maps.
\item Neuron Selectivity Transfer (NST)\cite{Huang2017NST}, which treats knowledge transfer as a distribution matching problem by aligning neuron selectivity patterns between networks.
\item Similarity-Preserving (SP)\cite{Tung2019SP}, which ensures the preservation of pairwise similarities between activations in the teacher network to maintain relational integrity in the student network.
\item Relational Knowledge Distillation (RKD)\cite{Park2019RKD}, which focuses on transferring mutual relations of data samples using distance-wise and angle-wise distillation losses.
\item Probabilistic Knowledge Transfer (PKT)\cite{Passalis2018PKT}, which aligns the probability distributions in the feature space, rather than directly mapping the features.
\item Correlation Congruence (CC)\cite{Peng2019CC}, which transfers correlations between instances using a generalized kernel method based on the Taylor series expansion.
\end{enumerate}

The classification accuracies (\%) of the baseline methods and the proposed ResizeNet+ approaches are summarized in Table~\ref{tab:kaggle_acc}.

\setcounter{table}{0}
\renewcommand{\thetable}{A\arabic{table}}
\begin{table*}[htbp] \centering \setlength{\tabcolsep}{0.5mm} \footnotesize
\caption{Average classification accuracies (\%) of unsupervised and semi-supervised cross-species transfer on Kaggle dataset. The best average accuracy of each task is marked in bold, and the second best by an underline.} \label{tab:kaggle_acc}
\begin{threeparttable}
\begin{tabular*}{\textwidth}{@{\extracolsep\fill}c|c|llllll|llllll}
\toprule
\multicolumn{2}{@{}c|@{}}{\multirow{2}{*}{Approach}}& \multicolumn{6}{c|}{Canine-to-Human} & \multicolumn{6}{c}{Human-to-Canine} \\\cmidrule{3-8}\cmidrule{9-14}%
\multicolumn{2}{@{}c|@{}}{}& 0\% & 5\% & 10\% & 15\% & 20\% & Avg. & 0\% & 5\% & 10\% & 15\% & 20\% & Avg. \\
\midrule
\multirow{7}{*}{Baseline}
& Within & / & 81.77 & 83.13 & 86.31 & 86.72  & 84.48  & / & 85.33 & 87.61 & 90.39 & 91.73 & 88.77 \\
& Comb. & 79.21 & 88.52 & 89.17 & 89.20 & 89.86 & 87.19  & 88.21 & 90.31 & 90.01 & 90.29 & 89.76 & 89.72 \\
& DAN & 85.11*** & 87.14 & 88.58 & 89.29 & 90.41 & 88.11  & 85.25 & 90.48 & 91.46* & 92.90** & \underline{94.17}** & 90.85 \\
& JAN & 85.03**** & 86.32 & 87.33 & 88.40 & 89.66 & 87.35  & 85.11 & 89.41 & 91.17* & 93.10*** & 93.22* & 90.40 \\
& SHOT & 72.74 & 77.00 & 78.92 & 80.28 & 81.90 & 78.17  & 75.60 & 80.90 & 84.84 & 85.79 & 86.54 & 82.73 \\
& DSAN & 82.89* & 86.27 & 86.91 & 88.66 & 87.52 & 86.45  & 86.45 & 90.58 & \underline{92.34}** & 93.30 & 94.16** & 91.37 \\
& MCC & 75.89 & 87.34 & 89.54 & 89.75 & 91.59 & 86.82  & 80.08 & \underline{92.03} & 92.11** & \underline{93.84}*** & 93.85** & 90.38 \\
\midrule
\multirow{7}{*}{ResizeNet+}
& AT & 82.92 & 89.56* & 90.88** & 92.14* & \underline{93.58}** & 89.82  & \underline{89.28}* & 90.59 & 91.08 & 92.68* & 93.66** & 91.46 \\
& NST & 85.02*** & 87.49 & 88.23 & 89.01 & 89.70 & 87.89  & 88.35 & 91.39 & 91.66 & 91.38 & 91.40* & 90.84 \\
& SP & 84.38** & 89.32 & \textbf{91.27}* & \underline{92.24} & 94.06** & 90.25  & 89.10 & 90.75 & 91.91** & 92.59* & 92.89* & 91.45 \\
& RKD & 80.22 & 89.14 & 91.07** & 91.20 & 92.65** & 88.86  & 88.04 & 90.96 & 91.94 & 91.37 & 92.93 & 91.05 \\
& PKT & 83.04* & 89.54 & 90.41 & 91.77 & \textbf{93.65}** & 89.68  & 89.18 & 90.75 & 92.21** & 92.66* & 93.77** & \underline{91.71} \\
& CC & \underline{85.19}*** & \underline{89.84} & 91.53* & 92.24* & 93.46* & \underline{90.45}  & 88.35 & 91.07 & 91.30 & 91.83 & 92.68* & 91.05 \\
& MSA & \textbf{85.55}*** & \textbf{90.29}** & \underline{91.26}** & \textbf{92.34}* & 93.54** & \textbf{90.60}  & \textbf{89.58}* & \textbf{92.90}* & \textbf{93.72}*** & \textbf{94.80}*** & \textbf{95.05}*** & \textbf{93.21} \\
\bottomrule
\end{tabular*}
\begin{tablenotes}
\item {****: $p<0.0001$; ***: $p<0.001$; **: $p<0.01$; *: $p<0.05$.}
\end{tablenotes}
\end{threeparttable}
\end{table*}

\section{APPENDIX B: DATASET INFORMATION} \label{secA2}

The Kaggle dataset includes data from eight human patients and four canines, provided by the Mayo Clinic and University of Pennsylvania. The canines were continuously monitored via video and iEEG. The iEEG data were acquired from an implanted device with a sampling rate of 400 Hz, utilizing 16 subdural electrodes arranged on two standard, human-sized, 4-contact strips implanted in an anteroposterior position on each hemisphere. The remaining eight human patients had drug-resistant epilepsy and were undergoing iEEG monitoring at Mayo Clinic Rochester. These signals were continuously sampled at either 500 or 5000 Hz, using varying subdural electrode grids based on individual clinical considerations\cite{brain2017kaggle}.

The Freiburg dataset comprises invasive long-term iEEG recordings from 21 human patients, acquired at a sampling rate of 256 Hz during invasive pre-surgical epilepsy monitoring at the Epilepsy Center of the University Hospital of Freiburg, Germany. The iEEG signals were recorded with three focal and three extra-focal electrode contacts\cite{Freiburg2012}.

The CHSZ dataset consists of sEEG recordings from 27 children, aged three months to ten years. The original sampling rate was either 500 Hz or 1000 Hz. Each subject had one to six seizure events. Experts annotated the onset and offset of each seizure for every child\cite{wang2023tasa}.

The NICU dataset documented neonatal seizures from 79 full-term neonates, with a sampling frequency of 256 Hz and a median recording duration of 74 minutes (interquartile range: 64 to 96 minutes). Three experts independently annotated each second of sEEG data, yielding an average of 460 seizures per expert. According to their consensus, 39 neonates experienced seizures, and these data were incorporated into our experiments\cite{nicu2019}.

Table~\ref{dataset_info} shows the main characteristics of four datasets. Tables~\ref{tab:kaggle_info}-\ref{tab:chsz_info} provide detailed characteristics for each subject in the Kaggle, Freiburg and CHSZ datasets, respectively.

\setcounter{table}{0}
\renewcommand{\thetable}{B\arabic{table}}
\begin{table*}[h]
\small
\setlength{\tabcolsep}{1mm}
\renewcommand\arraystretch{1}
\caption{Summary of the four epilepsy datasets.}\label{dataset_info}
\begin{tabular*}{\textwidth}{@{\extracolsep\fill}llcccccc}
\toprule
\multirow{2}{*}{Dataset}
& \multirow{2}{*}{EEG Type} & \multirow{2}{*}{\# Patients}  & \multirow{2}{*}{\# Channels} & Sampling rate & Signal length & \# Ictal & \# Interictal\\
& & & & (Hz) & (second) & trials & trials\\
\midrule
\multirow{2}{*}{Kaggle}
& \multirow{2}{*}{Intracranial} & 4 canines  & 16 & 400 & 1 & 1,087 & 9,116\\
& & 8 humans & [16, 72] & 500 or 5000 & 1 & 1,390 & 14,329 \\
Freiburg & Intracranial  & 21 humans  &  6 & 256 & 1 & 21,000 & 189,000 \\
NICU & Scalp & 39 humans   & 18  & 256 & 4 & 11,912 & 40,622\\
CHSZ & Scalp & 27 humans   & 18 & 500 & 4 & 716 & 20,521 \\
\bottomrule
\end{tabular*}
\end{table*}

\setcounter{table}{1}
\renewcommand{\thetable}{B\arabic{table}}
\begin{table*}[htpb] \centering \setlength{\tabcolsep}{1mm}
\small
\caption{Characteristics of the four canine subjects and eight human subjects in the Kaggle dataset.}\label{tab:kaggle_info}
\begin{tabular*}{\textwidth}{@{\extracolsep\fill}c|ccccccc}
\toprule
Species & ID & \# Channels & Sampling rate (Hz) & \# Seizures & \# Ictal trials & \# Interictal trials & \# Total trials  \\
\midrule
\multirow{4}{*}{Canine}
& 1 & 16 & 400 & 9 & 178 & 418 & 596  \\
& 2 & 16 & 400 & 5 & 172 & 1,148 & 1,320  \\
& 3 & 16 & 400 & 22 & 480 & 4,760 & 5,240  \\
& 4 & 16 & 400 & 6 & 257 & 2,790 & 3,047  \\
\midrule
\multirow{8}{*}{Human}
& 1 & 68 & 500 & 7 & 70 & 104 & 174  \\
& 2 & 16 & 5,000 & 7 & 151 & 2,990 & 3,141  \\
& 3 & 55 & 5,000 & 9 & 327 & 714 & 1,041  \\
& 4 & 72 & 5,000 & 5 & 20 & 190 & 210  \\
& 5 & 64 & 5,000 & 7 & 135 & 2,610 & 2,745  \\
& 6 & 30 & 5,000 & 8 & 225 & 2,772 & 2,997  \\
& 7 & 36 & 5,000 & 6 & 282 & 3,239 & 3,521  \\
& 8 & 16 & 5,000 & 4 & 180 & 1,710 & 1,890 \\
\bottomrule
\end{tabular*}
\end{table*}

\setcounter{table}{2}
\renewcommand{\thetable}{B\arabic{table}}
\begin{table*}[htpb]  \centering \setlength{\tabcolsep}{1mm}
\small
\caption{Characteristics of the 21 human subjects in the Freiburg dataset.} \label{tab:freiburg_info}
\begin{threeparttable}
\begin{tabular*}{\textwidth}{@{\extracolsep\fill}ccccccccc}
\toprule
ID & Sex & Age & H/NC\tnote{1} & Origin of seizure & \# Seizures & \# Ictal trials & \# Interictal trials & \# Total trials \\
\midrule
1 & f & 15 & NC & Frontal & 4 & 25,200 & 86,400 & 111,600  \\
2 & m & 38 & H & Temporal & 3 & 20,243 & 86,400 & 106,643 \\
3 & m & 14 & NC & Frontal & 5 & 29,480 & 86,400 & 115,880  \\
4 & f & 26 & H & Temporal & 5 & 36,000 & 86,400 & 122,400  \\
5 & f & 16 & NC & Frontal & 5 & 35,630 & 86,400 & 122,030  \\
6 & f & 31 & H & Temporo/Occipital & 3 & 23,100 & 86,400 & 109,500  \\
7 & f & 42 & H & Temporal & 3 & 21,600 & 88,597 & 110,197  \\
8 & f & 32 & NC & Frontal & 2 & 12,871 & 86,979 & 99,850  \\
9 & m & 44 & NC & Temporo/Occipital & 5 & 36,000 & 86,163 & 122,163  \\
10 & m & 47 & H & Temporal & 5 & 38,535 & 88,047 & 126,582  \\
11 & f & 10 & NC & Parietal & 4 & 28,800 & 86,570 & 115,370  \\
12 & f & 42 & H & Temporal & 4 & 28,800 & 178,652 & 207,452  \\
13 & f & 22 & H & Temporo/Occipital & 2 & 14,400 & 86,400 & 100,800  \\
14 & f & 41 & H, NC & Fronto/Temporal & 4 & 25,200 & 85,894 & 111,094  \\
15 & m & 31 & H, NC & Temporal & 4 & 36,000 & 86,400 & 122,400  \\
16 & f & 50 & H & Temporal & 5 & 42,075 & 86,400 & 128,475  \\
17 & m & 28 & NC & Temporal & 5 & 53,285 & 86,634 & 139,919  \\
18 & f & 25 & NC & Frontal & 5 & 46,698 & 89,569 & 136,267  \\
19 & f & 28 & NC & Frontal & 4 & 46,800 & 87,780 & 134,580  \\
20 & m & 33 & NC & Temporo/Parietal & 5 & 46,472 & 92,219 & 138,691  \\
21 & m & 13 & NC & Temporal & 5 & 43,200 & 86,177 & 129,377  \\
\bottomrule
\end{tabular*}
\begin{tablenotes}
\item[1]{H: Hippocampal origin; NC: Neocortical origin.}
\end{tablenotes}
\end{threeparttable}
\end{table*}

\setcounter{table}{3}
\renewcommand{\thetable}{B\arabic{table}}
\begin{table*}[htbp] \centering \setlength{\tabcolsep}{1mm}
\small
\caption{Characteristics of the 27 subjects in the CHSZ dataset.}
\label{tab:chsz_info}
\begin{threeparttable}
\begin{tabular*}{\textwidth}{@{\extracolsep\fill}ccccccc}
\toprule
ID & Age & Seizure subtype\tnote{1} & \# Seizures & \# Ictal trials & \# Interictal trials & \# Total trials \\ \midrule
1 & 10m & Tonic-Clonic & 1 & 21 & 170 & 191 \\
2 & 7m & Tonic-Clonic & 1 & 13 & 28 & 41 \\
3 & 3m & Tonic-Clonic & 1 & 21 & 100 & 121 \\
4 & 2y1m & Tonic-Clonic & 3 & 11 & 51 & 62 \\
5 & 8y11m & Partial & 1 & 46 & 16 & 62 \\
6 & 4y7m & Tonic-Clonic & 1 & 11 & 133 & 144 \\
7 & 10y10m & Absence & 1 & 10 & 16 & 26 \\
8 & 1y8m & Tonic-Clonic & 1 & 24 & 13 & 37 \\
9 & 1y7m & Tonic-Clonic & 1 & 14 & 52 & 66 \\
10 & 1y8m & Partial & 1 & 50 & 3,549 & 3,599 \\
11 & 1y6m & Partial & 4 & 143 & 2,595 & 2,738 \\
12 & 2y11m & Partial & 1 & 16 & 3,583 & 3,599 \\
13 & 3m & Partial & 1 & 30 & 224 & 254 \\
14 & 3y7m & Absence & 6 & 9 & 2,560 & 2,569 \\
15 & 6y4m & Absence & 6 & 27 & 158 & 185 \\
16 & 3m & Partial & 1 & 33 & 9 & 42 \\
17 & 2y8m & Tonic-Clonic & 1 & 44 & 53 & 97 \\
18 & 3y11m & Absence & 3 & 12 & 600 & 612 \\
19 & 3m & Partial & 1 & 16 & 1,321 & 1,337 \\
20 & 6y3m & Absence & 6 & 16 & 1,031 & 1,047 \\
21 & 6m & Partial & 1 & 23 & 82 & 105 \\
22 & 8y10m & Tonic-Clonic & 2 & 30 & 36 & 66 \\
23 & 8y6m & Partial & 4 & 8 & 77 & 85 \\
24 & 3y10m & Tonic-Clonic & 2 & 33 & 64 & 97 \\
25 & 4y6m & Tonic-Clonic & 2 & 29 & 89 & 118 \\
26 & 7y8m & Partial & 1 & 10 & 328 & 338 \\
27 & 5m & Partial & 1 & 16 & 3,583 & 3,599 \\
\bottomrule
\end{tabular*}
\begin{tablenotes}
\item[1]{An expert annotated the beginning and end of each seizure and its subtype (tonic-clonic seizure, absence seizure, or partial seizure) for each child.}
\end{tablenotes}
\end{threeparttable}
\end{table*}

\section{APPENDIX C: DATA PREPROCESSING}\label{secA3}

For the NICU and CHSZ datasets, the original sEEG signals were acquired using 19 unipolar electrodes positioned according to the international 10–20 system, from which 18 bipolar channels were derived\cite{wang2023tasa}: Fp2-F4, F4-C4, C4-P4, P4-O2, Fp1-F3, F3-C3, C3-P3, P3-O1, Fp2-F8, F8-T4, T4-T6, T6-O2, Fp1-F7, F7-T3, T3-T5, T5-O1, Fz-Cz, and Cz-Pz.

The EEG signals in all four datasets entered a 50 Hz notch filter and a 0.5-50 Hz bandpass filter, and were then segmented into 1-second non-overlapping trials in the Kaggle and Freiburg datasets, and 4-second trials in the CHSZ and NICU datasets, in line with\cite{wang2023tasa}.

In the NICU dataset, EEG segments were independently annotated by three experts on a per-second basis, with groundtruth labels determined by majority consensus. In the other three datasets, a single expert annotated the beginning and end of each seizure. For each cross-species transfer task, the duration of time segments was unified across the source and target species. For instance, when transferring from canines to humans in the Kaggle dataset, the human iEEG signals were downsampled to 400 Hz using the resample function in the MNE package (\url{https://mne.tools/stable/index.html}).

\section{APPENDIX D: CROSS-DATASET HUMAN-TO-HUMAN TRANSFER EXPERIMENTS}\label{secA4}

Within-species cross-dataset experiments were conducted to validate the effectiveness of the proposed ResizeNet+MSA approach.

There were significant between-dataset discrepancies in this setting. For example, the CHSZ dataset includes 18 sEEG channels sampled at 400 Hz, whereas the Kaggle-Human dataset consists of 16-72 iEEG channels sampled at 500 Hz. Although these datasets originate from the same species, they differ significantly in acquisition methods, channel numbers and locations, and sampling rates. Based on these variations, we explored three experiment settings, as illustrated in Figure~\ref{fig:cross_dataset_settings}:
\begin{enumerate}
\item Cross-dataset, within-species (human-to-human) and within-modality transfer: In this scenario, models were trained on human sEEG data from the NICU or CHSZ dataset and tested on another human sEEG dataset. Since there were no channel differences between the NICU and CHSZ datasets, the proposed ResizeNet+MSA approach was not applied in this setting.
\item Cross-dataset, within-species (human-to-human) and cross-modality transfer: Models were trained on human iEEG or sEEG datasets and tested on human datasets of a different modality (e.g., iEEG to sEEG, or sEEG to iEEG). Input discrepancies exist in this setting, making alignment necessary.
\item Cross-dataset, multi-species [(canine+human)-to-human] and multi-modality transfer: Canine iEEG and human iEEG/sEEG datasets were combined as the training set, with CHSZ used as the test dataset. This setting involves both multi-species and multi-modality data, introducing significant input discrepancies.
\end{enumerate}

\setcounter{figure}{0}
\renewcommand{\thefigure}{D\arabic{figure}}
\renewcommand{\dblfloatpagefraction}{.9}
\begin{figure*}[htpb]
\includegraphics[width=\linewidth,clip]{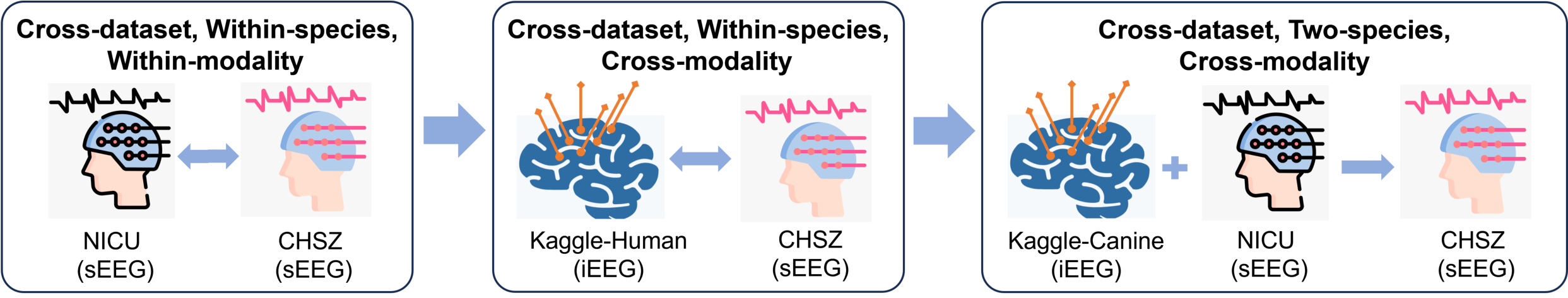}
\caption{Illustration of three cross-dataset transfer scenarios.} \label{fig:cross_dataset_settings}
\end{figure*}

\setcounter{table}{0}
\renewcommand{\thetable}{D\arabic{table}}
\begin{table*}[htbp] \centering \setlength{\tabcolsep}{0.5mm} \footnotesize
\caption{Average unsupervised cross-dataset transfer AUCs (\%) on NICU and CHSZ datasets. The best average performance of each task is marked in bold, and the second best by an underline.} \label{tab:wscdwm}
\begin{threeparttable}[b]
\begin{tabular*}{\textwidth}{@{\extracolsep\fill}c|c|cc|cc|cc}
\toprule
\multicolumn{2}{@{}c|@{}}{\multirow{8}{*}{Approach}}&
\multicolumn{2}{c|}{Within-species} & \multicolumn{2}{c|}{Within-species} & \multicolumn{2}{c}{Multi-species} \\
\multicolumn{2}{@{}c|@{}}{}&\multicolumn{2}{c|}{within-modality\tnote{1}} & \multicolumn{2}{c|}{cross-modality\tnote{2}} & \multicolumn{2}{c}{multi-modality\tnote{3}} \\ \cline{3-8}
\multicolumn{2}{@{}c|@{}}{}& NICU & CHSZ & Kaggle (H) & CHSZ & NICU+Kaggle (C) & NICU+Kaggle (C+H) \\
\multicolumn{2}{@{}c|@{}}{}&-to-&-to-&-to-&-to-&-to-&-to-\\
\multicolumn{2}{@{}c|@{}}{}& CHSZ & NICU & CHSZ & Kaggle (H) & CHSZ & CHSZ \\
\multicolumn{2}{@{}c|@{}}{}&(sEEG & (sEEG &(iEEG &(sEEG &(iEEG+sEEG &(iEEG+sEEG \\
\multicolumn{2}{@{}c|@{}}{}&-to- & -to- & -to- & -to- & -to- & -to- \\
\multicolumn{2}{@{}c|@{}}{}&sEEG) & sEEG)& sEEG)& iEEG) & sEEG)& sEEG) \\
\midrule
\multirow{6}{*}{Baseline\tnote{4}}
& Source Only\tnote{6} & \textbf{85.17}$_{\pm1.3}$ & 68.87$_{\pm1.0}$ & 84.37$_{\pm0.1}$ & 67.24$_{\pm1.3}$ &86.33$_{\pm0.69}$ & 87.05$_{\pm1.04}$\\
& DAN & 82.28$_{\pm2.1}$ & 68.84$_{\pm0.1}$ & 76.75$_{\pm2.8}$ & 72.66$_{\pm3.5}$ &82.57$_{\pm1.13}$ & 82.72$_{\pm1.69}$\\
& JAN & 80.81$_{\pm0.9}$ & 64.52$_{\pm0.7}$ & 70.01$_{\pm1.7}$ & \underline{73.79}$_{\pm1.9}$ &81.35$_{\pm1.86}$ & 80.10$_{\pm3.34}$\\
& SHOT  & 80.03$_{\pm1.1}$ & 68.82$_{\pm0.2}$ & 72.75$_{\pm3.5}$ & 53.92$_{\pm1.5}$ &74.44$_{\pm4.08}$ & 80.86$_{\pm1.00}$\\
& DSAN  & 68.97$_{\pm2.5}$ & \textbf{69.66}$_{\pm0.1}$ & 67.99$_{\pm1.8}$ & 66.51$_{\pm3.6}$ &60.09$_{\pm1.14}$ & 60.90$_{\pm1.21}$\\
& MCC  & \underline{83.71}$_{\pm0.3}$ & \underline{69.62}$_{\pm0.2}$ & 78.11$_{\pm4.3}$ & 56.05$_{\pm1.8}$ &75.27$_{\pm4.62}$ & 76.60$_{\pm5.40}$\\
\midrule
\multirow{7}{*}{ResizeNet+\tnote{5}}
& AT & -& -& 84.29$_{\pm1.4}$ & 67.03$_{\pm5.3}$ &89.14$_{\pm1.09}$ & 88.94$_{\pm0.77}$\\
& NST & -& -& \underline{84.87}$_{\pm1.7}$ & 70.11$_{\pm1.0}$ &89.07$_{\pm0.15}$ & \underline{89.46}$_{\pm0.28}$\\
& SP & -& -& 84.06$_{\pm1.1}$ & 67.05$_{\pm5.1}$ &89.19$_{\pm0.60}$ & 89.26$_{\pm0.40}$\\
& RKD & -& -& 80.40$_{\pm1.9}$ & 66.76$_{\pm0.9}$ &88.48$_{\pm0.55}$ & 89.21$_{\pm0.08}$\\
& PKT & -& -& 84.14$_{\pm1.1}$ & 67.11$_{\pm5.2}$ &88.84$_{\pm0.40}$ & 88.87$_{\pm0.46}$\\
& CC & -& -& 83.54$_{\pm1.2}$ & 69.26$_{\pm1.0}$ &\underline{89.46}$_{\pm1.39}$ & 88.64$_{\pm1.40}$\\
\cmidrule(r){2-8}
& MSA (ours) & -& -& \textbf{87.03}$_{\pm0.9}$ & \textbf{75.03}$_{\pm1.6}$ &\textbf{90.55}$_{\pm0.73}$ & \textbf{91.34}$_{\pm0.85}$\\
\bottomrule
\end{tabular*}
\begin{tablenotes}
\item[1]{Cross-dataset, within-species (human-to-human) and within-modality transfer: Train on human sEEG data from NICU or CHSZ dataset, and test on another human sEEG dataset. There are no channel differences between NICU and CHSZ datasets, thus the proposed ResizeNet+MSA approach was not performed in this setting.}
\item[2]{Cross-dataset, within-species (human-to-human) and cross-modality transfer: Train on human iEEG/sEEG dataset, and test on human dataset with another modality, i.e., iEEG to sEEG, or sEEG to iEEG. Input discrepancies exist in this setting.}
\item[3]{Cross-dataset, multi-species [(canine+human)-to-human] and multi-modality transfer: Canine iEEG and human iEEG/sEEG datasets were combined as the training set, and CHSZ was used as the test dataset. This setting used two species and two modality data to train the models, so that input discrepancies exist.}
\item[4]{Baseline: Without using the proposed ResizeNet, the number of channels for both species was unified by eliminating the mismatching ones.}
\item[5]{ResizeNet+: Utilize the proposed ResizeNet projection to unify the number of channels for different modalities.}
\item[6]{Source Only: Train the model on all labeled data from the other dataset without employing any alignment strategy.}
\end{tablenotes}
\end{threeparttable}
\end{table*}

Table~\ref{tab:wscdwm} shows the experiment results:
\begin{enumerate}
\item The cross-dataset within-species transfer performance was better than basic cross-species transfer without additional techniques.
\item ResizeNet+MSA demonstrated its effectiveness in cross-dataset cross-modality experiments.
\item Combining multi-species data in the training set and testing on a human dataset resulted in the best performance.
\item For the within-modality scenario, NICU and CHSZ datasets have the same number and placement of channels, so there were no data heterogeneous discrepancies. UDA approaches were not always effective. Compared to the cross-modality scenario, within-modality transfer achieved better performance, e.g., $85.17\%> 84.37\%$ and $68.87\%>67.24\%$, indicating that cross-modality transfer (with data heterogeneous discrepancy) is indeed more challenging.
\item For the cross-modality scenario, iEEG and sEEG signals, collected from different devices, introduce significant data heterogeneity. While UDA approaches may be effective in some cases, the proposed ResizeNet framework consistently outperformed others. Notably, ResizeNet+MSA achieved the highest performance among all approaches.
\item For the multi-modality scenario, we further combined two-species datasets to build the training set. With the increasing number of training data, the performance improved, e.g., $84.37\% < 86.33\% < 87.05\%$. With the help of ResizeNet+MSA, the best performance was achieved on the CHSZ dataset, e.g., 91.34\%, surpassing models trained on single-species single-modality data. This improvement likely stemmed from the increased training data volume and the proposed MSA, which effectively handles significant data heterogeneities.
\end{enumerate}

\section{APPENDIX E: SEIZURE PREDICTION PERFORMANCE}\label{secA4}

Accurate forecasting of epileptic seizures greatly improves clinical epilepsy diagnosis and patient care. However, significant interspecies differences pose challenges for seizure prediction tasks. To evaluate the effectiveness of the proposed ResizeNet+MSA approach, extensive experiments were conducted for cross-species seizure prediction.

The Kaggle seizure prediction dataset from the seizure forecasting competition \cite{Brain2016} was used, featuring epilepsy recordings from five canines and two humans undergoing extended iEEG monitoring. The data consist of 10-minute interictal and preictal clips. Substantial differences between canine and human recordings include sampling rates (400 Hz for canines, and 5000 Hz for humans) and electrode configurations (15 or 16 subdural electrodes for canines, and 15 or 24 intracranial channels for humans), reflecting cross-species characteristics similar to those in the seizure detection task. Additionally, the dataset exhibits a significant class imbalance between interictal and preictal trials. To address this imbalance, a subset of interictal samples was selected. For consistency with the seizure detection task, human iEEG signals were downsampled to 400 Hz, and each 10-minute clip was segmented into 10-second segments. Detailed characteristics of the dataset are summarized in Table~\ref{tab:dataset_info_prediction}.

Semi-supervised Canine-to-Human transfer experiments, with different portions of labeled trials, were conducted to evaluate three approaches: Within, Comb., and the proposed ResizeNet+MSA. Both Comb. and ResizeNet+MSA employed EA as a preprocessing step. The results, presented in Table~\ref{tab:aesp_result}, demonstrate that:
\begin{enumerate}
\item Comb. achieved superior performance compared to Within with an improvement of 6\%, highlighting the benefits of incorporating data from another species.
\item The proposed ResizeNet+MSA achieved an additional 4.6\% improvement over Comb., demonstrating its effectiveness in mitigating cross-species discrepancies and enabling better alignment across species.
\end{enumerate}

\setcounter{table}{0}
\renewcommand{\thetable}{E\arabic{table}}
\begin{table*}[htpb] \centering \setlength{\tabcolsep}{1mm}
\small
\caption{Characteristics of the five canine subjects and two human subjects in the Kaggle Prediction dataset.}\label{tab:dataset_info_prediction}
\begin{tabular*}{\textwidth}{@{\extracolsep\fill}c|ccccccc}
\toprule
Species & ID & \# Channels & Sampling rate (Hz) & \# Seizures & \# Preictal trials & \# Interictal trials & \# Total trials  \\
\midrule
\multirow{5}{*}{Canine}
& 1 & 16 & 400 & 22 & 1440 & 10,000 & 11,440 \\
& 2 & 16 & 400 & 47 & 2520 & 10,000 & 12,520  \\
& 3 & 16 & 400 & 104 & 4320 & 10,000 & 14,320  \\
& 4 & 16 & 400 & 29 & 5820 & 10,000 & 15,820  \\
& 5 & 15 & 400 & 19 & 1800 & 10,000 & 1,1800  \\
\midrule
\multirow{2}{*}{Human}
& 1 & 15 & 5,000 & 5 & 1,080 & 3,000 & 4,080  \\
& 2 & 24 & 5,000 & 41 & 1,080 & 2,520 & 3,600  \\
\bottomrule
\end{tabular*}
\end{table*}

\setcounter{table}{1}
\renewcommand{\thetable}{E\arabic{table}}
\begin{table*}[htpb] \centering \setlength{\tabcolsep}{2mm} \small
\caption{Average semi-supervised cross-species seizure prediction transfer AUC (\%) in Canine-to-Human seizure prediction. The best average performance of each task is marked in bold.}\label{tab:aesp_result}
\begin{tabular}{c|ccccc}
\toprule
\multirow{2}{*}{Approach} & \multicolumn{5}{c}{Portion of labeled trials} \\ \cline{2-6}
 & 5\% & 10\% & 15\% & 20\% & Avg. \\
\midrule
Within & 62.82$_{\pm0.9}$ & 72.81$_{\pm0.7}$ & 78.24$_{\pm2.2}$ & 82.61$_{\pm1.6}$ & 74.12\\
Comb. & 76.62$_{\pm1.0}$ & 82.11$_{\pm0.4}$ & 83.80$_{\pm1.4}$ & 84.59$_{\pm1.8}$ & 81.78\\
ResizeNet+MSA & \textbf{81.02}$_{\pm1.9}$ & \textbf{87.46}$_{\pm1.0}$ & \textbf{87.68}$_{\pm1.2}$ & \textbf{89.58}$_{\pm1.6}$ & \textbf{86.44}\\
\bottomrule
\end{tabular}
\end{table*}

\end{appendix}

\end{document}